\documentclass[pra,twocolumn,aps,amssymb,footinbib,showpacs]{revtex4}
\usepackage{amssymb}
\usepackage{graphicx}
\usepackage{amsmath}
\usepackage{times}
\usepackage{color}
\usepackage{subfigure}
\usepackage{setspace}
\usepackage{bm}

\begin{document}

\newcommand {\beq} {\begin{equation}}
\newcommand {\eeq} {\end{equation}}
\newcommand {\bqa} {\begin{eqnarray}}
\newcommand {\eqa} {\end{eqnarray}}
\newcommand {\no} {\nonumber}
\newcommand {\ep} {\ensuremath{\epsilon}}
\newcommand {\ms} {\medskip}
\newcommand {\bs} {\bigskip}
\newcommand{\rr} {\ensuremath{{\bf r}}}
\newcommand{\nbr} {\ensuremath{\langle ij \rangle}}
\newcommand{\nbrr} {\ensuremath{\langle kl \rangle}}
\newcommand{\ncap} {\ensuremath{\hat{n}}}
\newcommand{\can} {\ensuremath{{\cal S}}}
\newcommand{\vep}{\ensuremath{\varepsilon}}

\begin{abstract}
  We study the equilibrium and non-equilibrium properties of strongly
  interacting bosons on a lattice in presence of a random bounded
  disorder potential. Using a Gutzwiller projected variational
  technique, we study the equilibrium phase diagram of the disordered
  Bose Hubbard model and obtain the Mott insulator, Bose glass and
  superfluid phases. We also study the non equilibrium response of the
  system under a periodic temporal drive where, starting from the superfluid
  phase, the hopping parameter is ramped down linearly in time, and
  back to its initial value. We study the density of excitations
  created, the change in the superfluid order parameter and the energy
  pumped into the system in this process as a function of the inverse
  ramp rate $\tau$. For the clean case the density of excitations goes
  to a constant, while the order parameter and energy relaxes as
  $1/\tau$ and $1/\tau^2$ respectively. With disorder, the excitation
  density decays exponentially with $\tau$, with the decay rate
  increasing with the disorder, to an asymptotic value independent of
  the disorder.  The energy and change in order parameter also
  decrease as $\tau$ is increased.

\end{abstract}

\title{Quantum Dynamics of Disordered Bosons in an Optical Lattice }
\date{\today}
\author{  Chien-Hung Lin$^{1,2}$, Rajdeep Sensarma$^1$, K. Sengupta$^3$ and S. Das Sarma$^{1,2}$}
 \affiliation{ 1.~~Condensed Matter Theory Center, Department of Physics, University of Maryland, College Park, USA 20742\\
2.~~Joint Quantum Institute, University of Maryland, College Park, USA 20742\\
3.~~ Theoretical Physics Department, Indian Association for the Cultivation of Science, Kolkata 700032, India}
\maketitle

\section{Introduction}

The model of bosons on a lattice interacting repulsively through a
local interaction in the background of a random one-body disorder
potential (or the disordered Bose Hubbard
model~\cite{Fisher:BH:PRB:88}) has been used as a paradigm for
superfluid insulator transition in a host of disordered quantum
systems. These encompass a number of different condensed matter
systems, from $^4He$ on disordered
substrates~\cite{Reppy:He:Expt:PRB:97} or in porous media~\cite{Chan:He:Expt:PRL:03},  to dirty superconducting films~\cite{Goldman:SCfilm:Expt:PRL:91} and Josephson junction arrays~\cite{Mooij:JJ:Expt:PRB:96},
to disordered quantum magnets~\cite{Roscilde:QM:BG:Expt:arx:12}. In fact, this
model has often been used to describe the relevant bosonic degrees of
freedom near phase transitions in strongly disordered systems. There
are three main ingredients in this model: a hopping or kinetic energy
term for bosons, which tend to favour delocalized superfluid phases,
an onsite repulsion which tries to localize the bosons to create a
Mott insulator, and an onsite one-body random disorder potential which
scatters the bosons and lead to loss of coherence of the
superfluid. The interplay of these three different terms produces a
rich phenomenology in these systems, both in its equilibrium
properties and in terms of non equilibrium dynamics in these systems.

Beyond the traditionally material based phenomena, for which it serves
as a paradigm, the disordered Bose Hubbard model can be realized with
ultracold atomic
systems~\cite{Ignuscio:DBHEXPT:NAT:08,DeMarco:DBHEXPT:PRL:09,Aspect:DBHEXPT:PRL:10},
which have emerged as the new platform to study the behaviour of model
many-body Hamiltonians used in condensed matter physics and
elsewhere~\cite{Bloch:RMP:08}. The easy tunability of implemented
Hamiltonian parameters and almost complete isolation from external
environment makes these systems attractive candidates to simulate
strongly interacting quantum many body Hamiltonians, both on the
lattice and in the continuum. Although cold-atomic systems on optical
lattices are generally free of disorder (which is inevitably present
in solid state systems), disorder can be added in a controlled manner
either by use of speckle potentials~\cite{DeMarco:DBHEXPT:PRL:09,Aspect:DBHEXPT:PRL:10} or by the use of multiple optical
lattice beams with incommensurate wavelengths~\cite{Ignuscio:DBHEXPT:NAT:08,Biddle:PRA:09}. In either case, the
disorder potential (or its distribution in the case of speckle
potentials) is well characterized and the parameters characterizing
the disorder potential can be changed in a controlled way, in contrast
to condensed matter systems, where the disorder parameters are unknown
a-priori, and are mostly determined through a post-hoc process of
matching experimentally measured quantities (like transport
co-efficients) to theoretical model calculations. The possibility of
controlled addition of disorder, thus, makes cold atoms uniquely
suited to study the effects of disorder on strongly interacting
quantum many-body systems.

Cold atom systems also provide an added advantage of easy access to
the internal nonequilibrium dynamics of isolated interacting systems. The low
energy scales (in the absolute sense), the easy tunability of the
Hamiltonian parameters and the almost complete isolation of the system
from external environment make it very easy to perturb the system from
its equilibrium state in a well characterized way and then follow the
dynamics of the system without the help of ultrafast probes. This has
opened up the possibility of studying the quantum dynamics of these
systems out of equilibrium~\cite{Weiss:NEQBM:NAT:06,Esslinger:DBLDCY:PRL:10}.

Since the early work of Fisher {\it et al}~\cite{Fisher:BH:PRB:88},
the equilibrium properties of the disordered Bose Hubbard model has
been treated with various levels of sophistication from mean field
theory~\cite{Blakie:DBHMFT:PRA:07,Krutitsky:NJP:06,Pisarski:PRA:11} to strong coupling expansions~\cite{Freericks:PRB:96} to Monte Carlo
techniques~\cite{Pollet:MCDBH:PRL:09,Prokofyev:MCDBH:PRB:09,Ceperley:BG:PRB:11}.
In this paper, we provide an alternative approach to studying the
disordered Bose Hubbard model based on variational wavefunctions. Our
approach is applicable in the strongly interacting limit of the model,
but does not place any constraint on the strength of disorder
potential. The variational approach uses a canonical transformation to
systematically eliminate processes connecting states with large energy
difference ($\sim U$, the onsite Hubbard repulsion, or more) and
generates an effective low energy Hamiltonian for the system in the
strongly interacting limit. This effective Hamiltonian is then treated
with a Gutzwiller mean field wavefunction. There are several benefits
to this approach over other standard approaches : i) It captures the
strong correlations generated by the boson repulsion more accurately
than mean field theory ii) The requirement of disorder average makes
the problem numerically very resource intensive to treat beyond mean
field theory.  Our semi-analytic approach lessens the numerical
burden, while keeping essential ``beyond mean field''
correlations. iii) Since this approach generates an effective low
energy Hamiltonian, it can be easily modified to study quantum
dynamics in these systems. This is a crucial aspect of this approach,
which makes it qualitatively different from more sophisticated Monte
Carlo techniques, specially in larger than one dimensions.

In this paper, we first study the equilibrium phase diagram of the 2D
disordered Bose Hubbard model on a square lattice within our approach
as a function of $U/J$ and the chemical potential $\mu$ for different
values of the disorder strength $V$. This yields three phases: (a) an
incompressible phase incoherent Mott insulating phase at large
interaction strength, whose area decreases with increasing disorder
strength (b) a superfluid phase, with coherent condensation of the
bosons into a single quantum state at small interaction strength, and
(c) a Bose glass phase in between them, where the system is
compressible, but the phase coherence of the bosonic condensate is
completely destroyed. We also study the non-equilibrium dynamics of
the system under the following conditions: the system is initialized
in its ground state in the superfluid phase. The interaction parameter
is ramped up linearly in time to a very high value and then ramped
back linearly to its initial value. At the end of this process, we
study the density of excitations produced in the system, the energy
pumped into the system and the deviation of the superfluid order
parameter from its initial value, as a function of the rate of the
ramp, $1/\tau$. In the clean case, the excitation density goes to a
constant, while the order parameter deviation and energy scales as
$1/\tau$ and $1/\tau^2$ in the large $\tau$ limit. With disorder, the
excitation density shows an exponential decay. The energy and order
parameter deviation also decreases with increasing $\tau$, although a
scaling form is hard to obtain due to inherent noise in the data.

The paper is organized as follows: In section ~\ref{vwf}, we
present our variational wavefunction approach and introduce the
canonical transformation. Section ~\ref{cant} presents the details
of obtaining the canonical transformation operator and the effective
low energy Hamiltonian. In section ~\ref{eqpd}, we present the
equilibrium phase diagram calculated within our approach. In section
~\ref{neqd} we present the results for the non equilibrium dynamics
in the system. Finally, we conclude in section ~\ref{concl} with a
summary of our results and a discussion of limitations of the
present formalism and ways to improve it.

\section{Variational Wavefunction\label{vwf}}

The Hamiltonian of the disordered Bose Hubbard model on a square
lattice is given by
\beq
H=-J\sum_{\nbr}b^\dagger_ib_j+
\frac{U}{2}\sum_i\ncap_i(\ncap_i-1) +\sum_i (v_i-\mu)\ncap_i
\eeq
where $b^\dagger_i$ creates a boson on site $i$ and $\ncap$ is the boson
number operator. Here $J$ is the nearest neighbour hopping energy
scale, $U$ the on-site Hubbard repulsion, $\mu$ the chemical potential
and $v_i$ is the random local potential on site $i$. $v_i$ is a
spatially uncorrelated random variable drawn from a uniform
distribution in the range $-V/2 < v_i< V/2$, where $V$ sets the energy
scale for disorder effects.

The clean Bose Hubbard model ($v_i=0$) has a quantum phase transition
between a strongly interacting incompressible Mott insulating phase
with commensurate integer filling at small $J/U$ and a phase coherent
superfluid state with large number fluctuations at large $J/U$. The
transition is characterized by the vanishing of both the superfluid
stiffness and the compressibility as one reaches the Mott phase. In
the presence of disorder, there is an intervening Bose glass phase
where the superfluid stiffness vanishes, but the compressibility
remains finite.

We wish to study the equilibrium phases and dynamics in the
disordered Bose Hubbard model through a variational wavefunction
approach. For the equilibrium phase diagram at $T=0$, we use a
variational ground state wavefunction of the form
\beq |\psi \rangle = e^{-i\can}|\psi_0\rangle~~~~
|\psi_0\rangle=\prod_i \sum_n f_{ni}|n\rangle_i.
\eeq
Here $|\psi_0\rangle$ is a Gutzwiller type local mean field state with
variational parameters $f_{ni}$, which satisfies $\sum_n |f_{ni}|^2=1$
to ensure normalization of the state, $|n\rangle_i$ is the number
state with $n$ bosons on site $i$, and $e^{-i\can}$ is a canonical
transformation that builds in non-local correlations in the proximity
of a Mott insulator.

The canonical transformation approach has a long history of use in
the context of Fermi Hubbard model in the strongly interacting
limit, where it is used to convert the Hubbard model to the so
called ``t-J'' model used in the study of high temperature
superconductors~\cite{Girvin:CANTFH:PRB:88}. Recently this approach
has been adapted successfully to study the equilibrium phases of and
quantum dynamics in clean Bose Hubbard
model~\cite{Trefzger:DYN:PRL:11,Krish:11}. The canonical transform uses the
local number states (which are eigenstates of the local part of the
Hamiltonian) as the starting point. Note that in the present
formulation of the canonical transformation, we do not use a
particular local state as our starting point (except assuming a
local number state), as is done in Ref.
~\onlinecite{Trefzger:DYN:PRL:11}, where the atomic limit Mott phase
ground state with the same number of particles on each site is used
as the starting point. The hopping terms then start to build in
correlations between different number states on neighbouring sites.

The hopping term can connect local number states which differ in
energy by $\sim U$ or higher. To see this consider a state with $n_1$
particles on site $i$ and $n_2$ particles on site $j$ and a hopping
process where a particle hops from $j$ to $i$. The energy difference
(coming from the local part of the Hamiltonian) between the initial
and final state is $\delta \epsilon = U(n_1-n_2+1)+v_i-v_j$ and
$|\delta\epsilon|$ can be $\sim U$ or more depending on $n_1$ and
$n_2$. We would like to note that, since we are interested in energy
difference of states connected by hopping (which does not change the
total number of particles in the system), the chemical potential drops
out of the expression for the canonical transform. Hence our formalism
is applicable for any $\mu$, even to the parameter regime where
$\mu\sim nU$, $n$ being an integer.

The basic idea of the canonical transform is to eliminate terms in the
Hamiltonian which connects local number states differing by a large
energy ($\sim U$ or more) order by order in $J/U$ through the
canonical transformation. The easiest way to see this is to note that
for any operator $A$,
\beq
\langle \psi|A|\psi\rangle=\langle
\psi_0|A^\ast|\psi_0\rangle,~~~ where~~~ A^\ast=e^{i\can}Ae^{-i\can}
\eeq
is the canonically transformed operator. For the Hamiltonian, $H$, the
requirement that $H^\ast$ does not have any terms connecting states
which differ in energy by $\sim U$ fixes the form for $\can$. The low energy effective Hamiltonian, $H^\ast$, obtained by the canonical
transform, not only allows the low energy hopping processes, but also builds in correlations from virtual transitions to high energy states.
In the next section, we provide the details of the canonical
transformation and the effective Hamiltonian for the disordered Bose
Hubbard model.  We would like to note here that although we will focus
here on a random disorder potential, our formalism is capable of
handling \emph{any} one-body potential, e.g.it can be used to treat effects
of harmonic traps in ultracold atomic gases in optical lattice.

\section{The Canonical Transformation\label{cant}}

The disordered Hubbard model can be separated into a local part
containing the interaction and the one body potential and a kinetic
energy part.
\beq
H=H_0+\sum_{\nbr}T_{ij}, ~~~ H_0=\frac{U}{2}\sum_i\ncap_i(\ncap_i-1)-\mu_i\ncap_i
\eeq
where $\mu_i=\mu-v_i$, $\mu$ being the chemical potential and $v_i$
the random disorder potential. It is also easy to see that the hopping
term $T_{ij}=-Jb^\dagger_ib_j$ connects local states differing in energy
by $\vep^\alpha_{ij}=\alpha U+v_i-v_j$, where $\alpha= 0,\pm 1,\pm 2
..$. This suggests breaking up the hopping term,
$T_{ij}=\sum_{\alpha}T^\alpha_{ij}$, where
\bqa
 T^\alpha_{ij}&=&-Jb^\dagger_ib_j\delta(n_i-n_j-\alpha+1)\\
\no & =&-J\sum_ng^n_\alpha |n+1\rangle_i|n-\alpha\rangle_j \left._i\langle n|\left._j\langle n-\alpha+1|\right.\right.
\eqa
where $g^n_\alpha=\sqrt{(n+1)(n-\alpha+1)}$.  Here $T^\alpha_{ij}$
connects states with energy difference $\vep^\alpha_{ij}$. A mathematical way
of representing this information is the identity
\beq
[H_0,T^\alpha_{ij}]=\vep^\alpha_{ij}T^\alpha_{ij},
\label{eq:comm}
\eeq
which will be useful later in deriving the canonical transformation operator.

For weak disorder ($V \ll U$), it is evident that $T^0_{ij}$
represents a low energy hopping process, while $T^\alpha_{ij}$ for
$\alpha\neq 0$ changes energy of the state by an amount $\sim \alpha
U$ and has to be eliminated by the canonical transform. This breakup
of the kinetic energy term follows the method of Girvin {\it et al}~\cite{Girvin:CANTFH:PRB:88}
for fermionic Hubbard model with one crucial difference: in the Fermi
Hubbard model, the local Hilbert space is constrained by Pauli
exclusion and hence $\alpha =0,\pm 1$, whereas in the Bosonic model,
the infinite Hilbert space leads to $\alpha$ taking all possible
integer values. In practice, the local Hilbert space is cutoff at some
high value of occupancy number, and $\alpha$ will be restricted
accordingly.  This formalism can be generalized to strong disorder
potentials with some more complications, which will be discussed in a
future work.

The canonical transformation operator $i\can$ has an expansion in
$J/U$, i.e.  $i\can =i\can^1 +i\can^2 +....$, where $i\can^m \sim
(J/U)^m$ and terms upto $i\can^m $ completely removes high energy
terms upto order $J(J/U)^{m-1}$. Using the identity,
eqn.~(\ref{eq:comm}), it can be shown that
\bqa
i\can^1&=& \sum_{\nbr}\sum_{\alpha\neq 0}\frac{T^\alpha_{ij}}{\vep^\alpha_{ij}}
\eqa
removes all high energy terms ${\cal O} (J)$, while
\bqa
 i\can^2&=&\sum_{\nbr\nbrr}\sum_{\alpha\neq 0}\frac{[T^\alpha_{ij},T^0_{kl}]}{\vep^\alpha_{ij}(\vep^\alpha_{ij}+v_k-v_l)}\\
\no & & + \frac{1}{4}\sum_{\nbr\nbrr}\sum_{\alpha \neq \beta\neq 0}\frac{[T^\alpha_{ij},T^{-\beta}_{kl}]}{(\vep^{-\beta}_{kl}+\vep^\alpha_{ij})}\left[\frac{1}{\vep^\alpha_{ij}}-\frac{1}{\vep^{-\beta}_{kl}}\right]
\eqa
removes high energy terms upto ${\cal O}(J^2/U)$.
\begin{figure}[t]
\centering
\includegraphics[width=0.75\columnwidth]{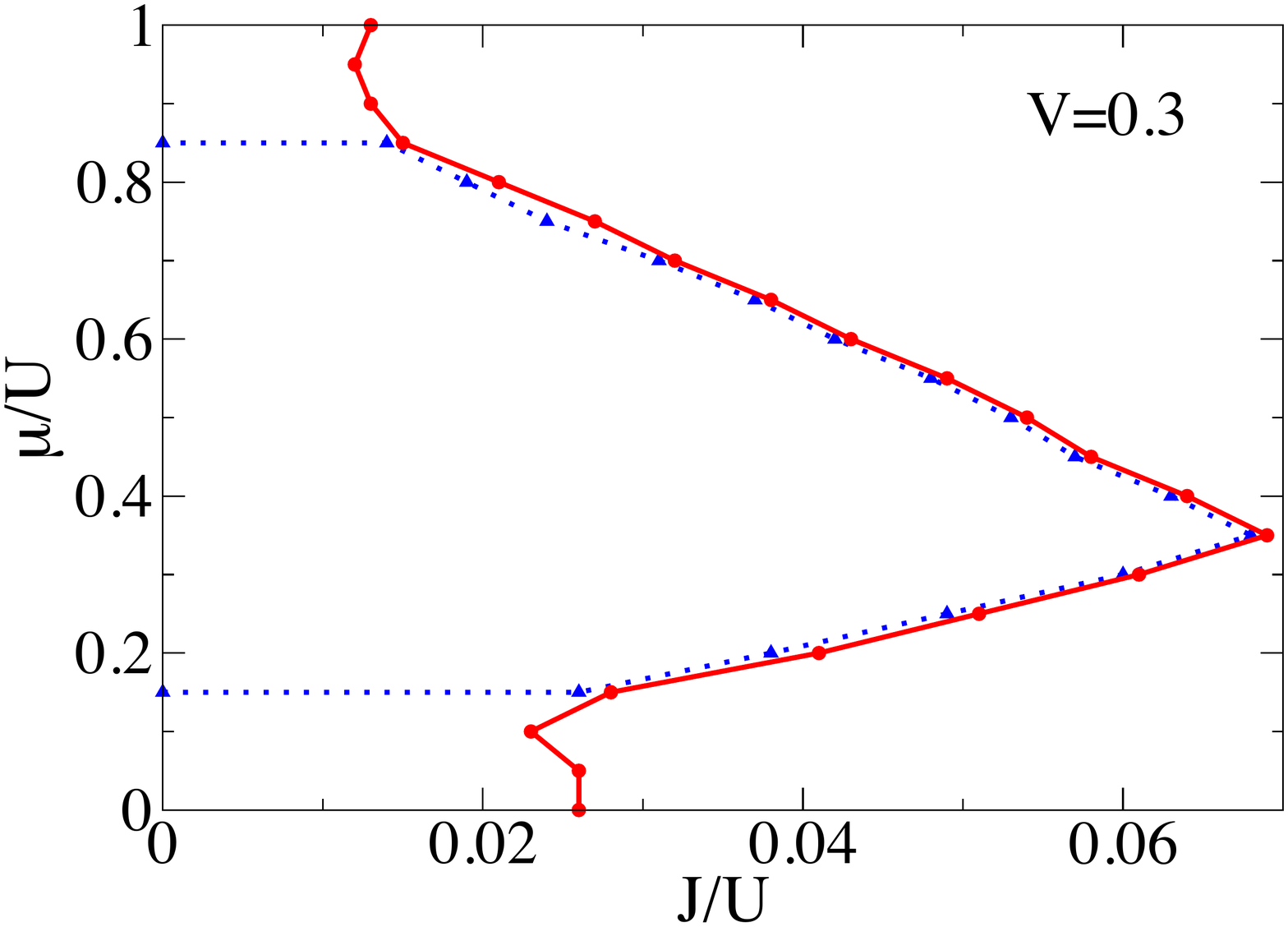}
\includegraphics[width=0.75\columnwidth]{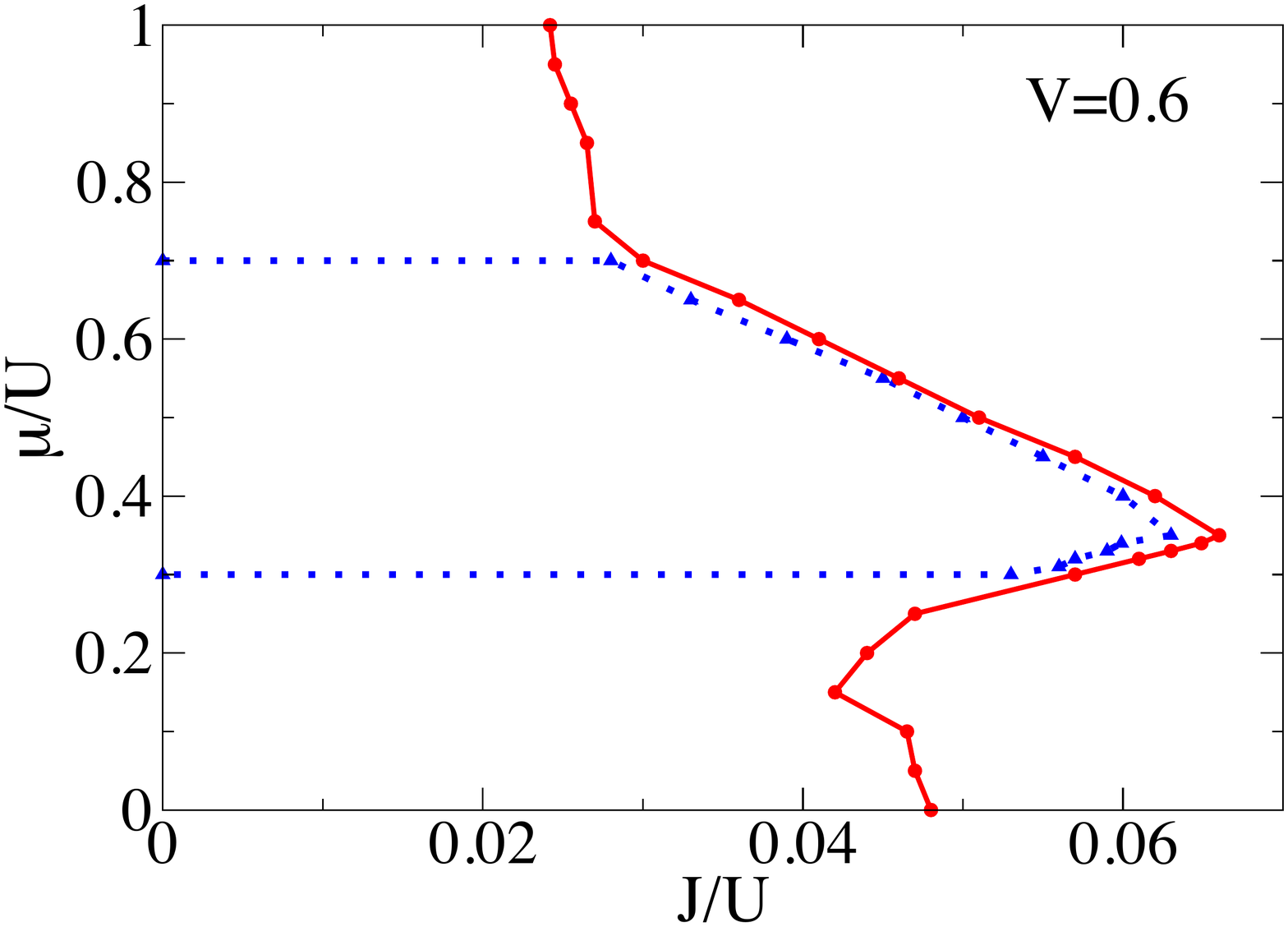}
\caption{ The zero temperature equilibrium phase diagram of the
disordered Bose Hubbard model in the $\mu/U$-$J/U$ plane for (a)
$V/U=0.3$ and (b) $V/U=0.6$ respectively. The phase to the right of
the thick red line is the superfluid phase, while the Mott phase is
enclosed by the dotted blue line. The phase in between is the Bose
glass phase.} \label{fig:phdiag}
\end{figure}

The effective Hamiltonian $H^\ast$ is then given by
\beq
H^\ast=H_0+\sum_{\nbr}T^0_{ij}+\frac{1}{2}\sum_{\nbr\nbrr}\sum_{\alpha\neq 0}\frac{[T^\alpha_{ij},T^{-\alpha}_{kl}]}{\vep^\alpha_{ij}}
\eeq
which, in the clean case of a Bose Hubbard model without disorder, reduces to
\beq
H^\ast(V=0)=H_0+\sum_{\nbr}T^0_{ij}+\frac{1}{2}\sum_{\nbr\nbrr}\sum_{\alpha\neq 0}\frac{[T^\alpha_{ij},T^{-\alpha}_{kl}]}{\alpha U}
\eeq
The effective low energy Hamiltonian thus consists of three terms :
(a) $H_0$ which gives the local interaction and disorder potential,
(b) $T_0$, which represents the low energy hopping, and (c) the last
commutator, which can be easily interpreted as a second order
perturbation, and takes care of virtual transitions to high energy
states.

We would like to note that our canonical transformation improves upon
previous formulation by Trefzger {\it et. al}~\cite{Trefzger:DYN:PRL:11,Krish:11} in the
following ways: i) It can handle non-uniform states with arbitrary
one-body potentials in the local part of the Hamiltonian, which is
crucial in treating disordered bosons and ii) It takes into account
the full Hilbert space for bosons and is not an expansion around a
state with a fixed number of particles on each site. This is crucial
to look at the Bose glass phase (and to study properties of bosons in a trap), where the density varies from site
to site. This formulation can also handle more accurately the
superfluid phase near the Mott lobes, as it treats all the states in
the local Hilbert space on equal footing. In fact, in the clean case,
if one keeps only three states in the local Hilbert space (the
commensurate density in the Mott state, $n_0$ and $n_0\pm 1$), then
$\alpha$ is restricted to $0, \pm 1$, and our formulation reduces to
that in Ref.~\onlinecite{Trefzger:DYN:PRL:11}.

\section{Equilibrium Phase Diagram\label{eqpd}}

Starting from the early prediction of Fisher {\it
  et. al}~\cite{Fisher:BH:PRB:88}, the equilibrium phase diagram of
disordered Bose Hubbard model has been worked out by several previous
authors using various techniques ranging from mean field
theory~\cite{Blakie:DBHMFT:PRA:07} to quantum Monte Carlo
techniques~\cite{Prokofyev:MCDBH:PRB:09,Pollet:MCDBH:PRL:09}. Although
our main motivation is to study dynamics of the system when
interaction parameters are tuned, we present the equilibrium phase
diagram obtained by our method for the sake of completeness. This will
also set the stage for our study of dynamics in two ways: (i) the
equilibrium ground state forms the initial condition for the dynamics
of the system and (ii) we would like to know the trajectory of the
system, i.e. whether it goes into the Mott or the Bose glass phase as
we ramp up the interaction parameter starting from the superfluid
phase.

The ground state is obtained by minimizing the energy in the
variational state, which is equivalent to minimizing the expectation
of $H^\ast$ in the the mean field state $|\psi_0\rangle$. A
straightforward algebra shows that the ground state energy is a sum of six
different contributions, ${\cal E}=\sum_{r=0}^5{\cal E}_r$, where
\beq
{\cal E}_0=\sum_{ni}\left[\frac{U}{2}n(n-1)-\mu_in\right]|f_{ni}|^2
\eeq
is the local energy corresponding to the interaction and disorder potential,
\beq
{\cal E}_1=-J\sum_{n\nbr}(n+1)f^\ast_{n+1i}f_{ni}f^\ast_{nj}f_{n+1j}
\eeq
is the low energy nearest neighbor hopping,
\begin{widetext}
\beq
{\cal E}_2=\frac{J^2}{2}\sum_{\nbr}\sum_{n\alpha\neq 0}(n+1)|f_{ni}|^2\frac{(n+\alpha+1)|f_{n+\alpha+1j}|^2-(n-\alpha+1)|f_{n-\alpha+1j}|^2}{\vep^\alpha_{ij}}
\eeq
is the second order density-density interaction energy,
\beq
{\cal
  E}_3=\frac{J^2}{2}\sum_{\nbr}\sum_{n}(n+1)(n+2)f^\ast_{n+2i}f_{nj}f^\ast_{nj}f_{n+2j}\left[\frac{1}{\vep^1_{ij}}-\frac{1}{\vep^{-1}_{ij}}\right]
\eeq
is a second order pair hopping term where two bosons hop to the
nearest neighbor.
\beq
{\cal E}_4=\frac{J^2}{2}\sum_{\nbr\langle ik\rangle}\sum_{n\alpha\neq 0}\frac{f^\ast_{n+2i}f_{ni}}{\vep^\alpha_{ij}}[g^n_{-\alpha}g^{n+1}_\alpha f^\ast_{n+\alpha k}f_{n+\alpha+1k}f^\ast_{n-\alpha+1j}f_{n-\alpha+2j}-(\alpha\rightarrow -\alpha)]+h.c.
\eeq
which represents a second order process where two bosons from two different neighboring sites hop onto a site and its reverse process, and finally
\beq
{\cal E}_5=\frac{J^2}{2}\sum_{\nbr\langle jk\rangle}\sum_{n\alpha\neq 0}\frac{|f_{nj}|^2}{\vep^\alpha_{ij}}[g^n_{-\alpha}g^{n+\alpha}_\alpha f^\ast_{n+\alpha k}f_{n+\alpha+1k}f^\ast_{n+\alpha+1i}f_{n+\alpha i}-(\alpha\rightarrow -\alpha)]+h.c.
\eeq
which represents a second order next nearest neighbor hopping process.
\end{widetext}

The energy is then minimized with respect to the variational
parameters $f_{ni}$ to obtain the ground state wavefunction. The three
different phases are then identified according to the following
criterion: The superfluid phase is characterized by a non
vanishing superfluid stiffness, which controls the energy of the
system for long wavelength distortion of the phase of the Bose
Einstein condensate. This can alternatively be thought as the
diamagnetic response of the system to a vector potential. In presence
of a static vector potential $A$ along the $x$ direction, the hopping
parameters acquire an Ahronov-Bohm phase, $J_{ij} \rightarrow J_{ij}
e^{A(x_i-x_j)}$, and correspondingly $H \rightarrow H_A$ and $\can \rightarrow \can_A$. The superfluid stiffness can then be calculated as
\beq
\rho_s = \frac{1}{N_c}\sum_{{\cal C}}\frac{\partial^2 \langle H^\ast_A\rangle_0}{\partial A^2}\vert_{A=0}
\eeq
where ${\cal C}$ denotes disorder configurations and $N_c$ is the
number of configurations kept in the disorder average (typically $\sim
100$ in our calculations). We note that in case of finite disorder
potential we will always work with disorder averaged quantities in
this paper. Any state with $\rho_s\neq 0$ will be identified as a
superfluid phase. In the non-superfluid phase, we distinguish between
the Bose glass phase and the Mott insulating phase by the fact that
the Bose glass phase has a finite compressibility, while the Mott
insulating phase is incompressible. Within our formalism, this implies
that $f_{n_0i}=1$ for all the lattice sites in all the disorder
configurations in commensurate Mott insulator of filling $n_0$, while
in the Bose glass phase, $max(f_{n_0i}) <1$. We note here that
although $f_{n_0i}=1$ for all lattice sites in a Mott phase, the
canonical transform mixes in virtual number fluctuations in the ground
state wavefunction.

In Fig.~\ref{fig:phdiag} (a) and (b), we study the phase diagram of
the system in the $J/U$-$\mu/U$ plane, focusing in and around the
$n_0=1$ Mott plateau, for different disorder strengths $V/U=0.3$ and
$V/U=0.6$ respectively. The parameter regime to the right of the thick
red line represents the superfluid phase with a non-zero superfluid
stiffness ($\rho_s\neq 0$). The region enclosed to the left of the
blue dotted line is the incompressible Mott phase, while the region in
between these two lines represents the Bose glass phase with non zero
compressibility but zero superfluid stiffness.

The phase diagram qualitatively captures the basic physics of the
disordered Hubbard model. In the atomic limit, ($J=0$), the system
remains in the Mott phase as long as $V/2<\mu<U-V/2$. The local
Hamiltonian $H_0$ is then optimized by the configuration of one
particle on each site. On the other hand, for $\mu< V/2$, there are
sites where the local Hamiltonian is optimized by a hole, while for
$\mu>U-V/2$, there are sites where the local Hamiltonian is optimized
by double occupancy. Thus the state in this limit has number
fluctuations (and hence is compressible) while the local nature of the
fluctuations imply that superfluid stiffness is $0$. This state is
thus in the Bose glass phase. As $J/U$ is increased, the Mott phase
first gives rise to a narrow region of Bose glass phase, which then
gives way to the superfluid phase. In the region, where he atomic
limit ground state is a Bose glass, we see a direct transition between
a Bose glass and a superfluid phase.

With increasing disorder strength, we find two distinct feature of the
phase diagram: (a) The Mott region shrinks with increasing $V/U$ and
(b) The direct Bose glass to superfluid transition takes place at
larger values of $J/U$. The first one can be easily explained by
noting that with increasing $V/U$, the width of the Mott phase in the
atomic limit ($U-V/2>\mu>V/2$) decreases. The second feature is
explained by the fact that stronger disorder leads to stronger
scattering and hence larger values of $J/U$ is required to restore
phase coherence and hence superfluidity in the system.
\begin{figure}[t]
\includegraphics[width = 0.38\columnwidth,angle=270]{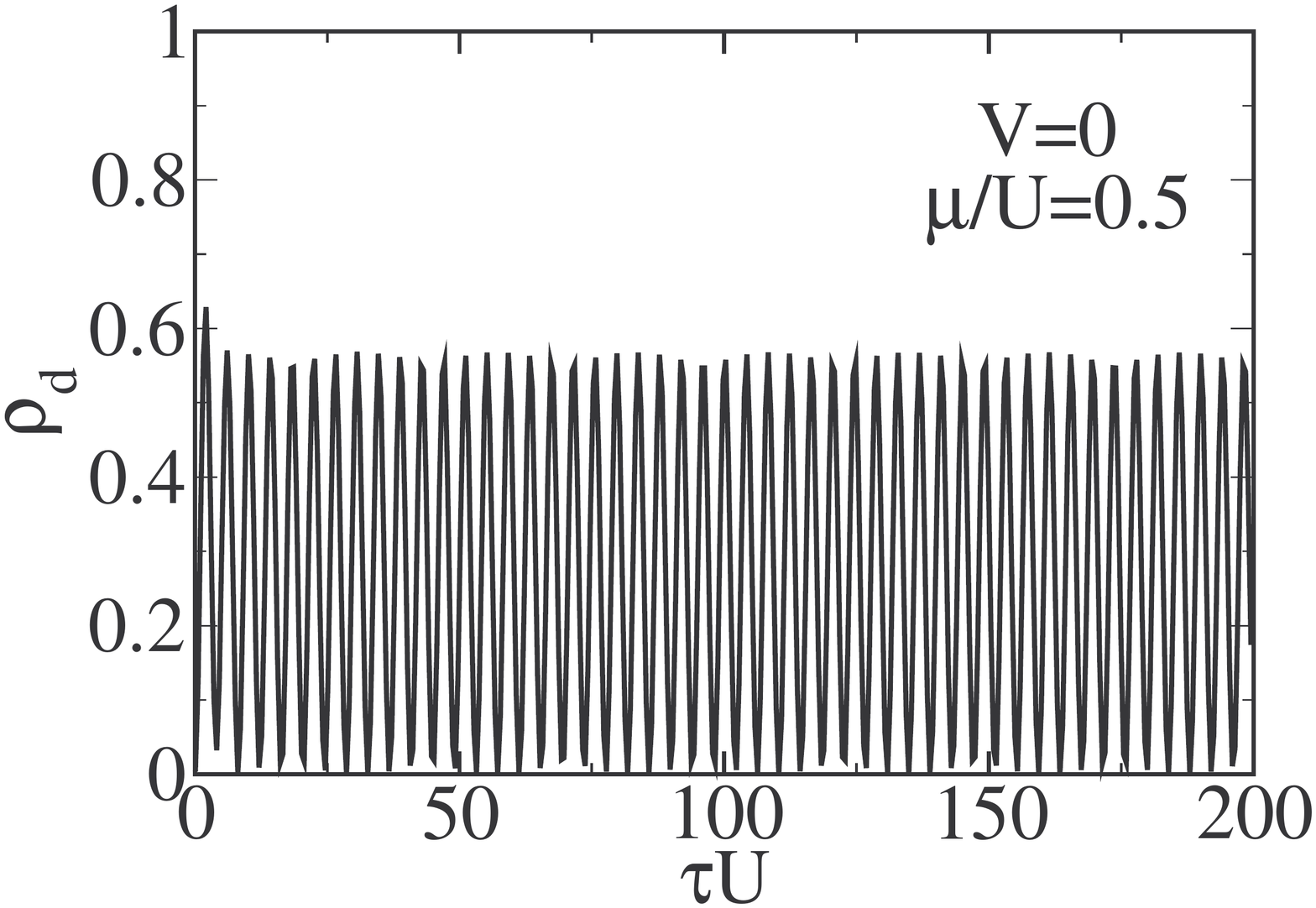}
\includegraphics[width = 0.38\columnwidth,angle=270]{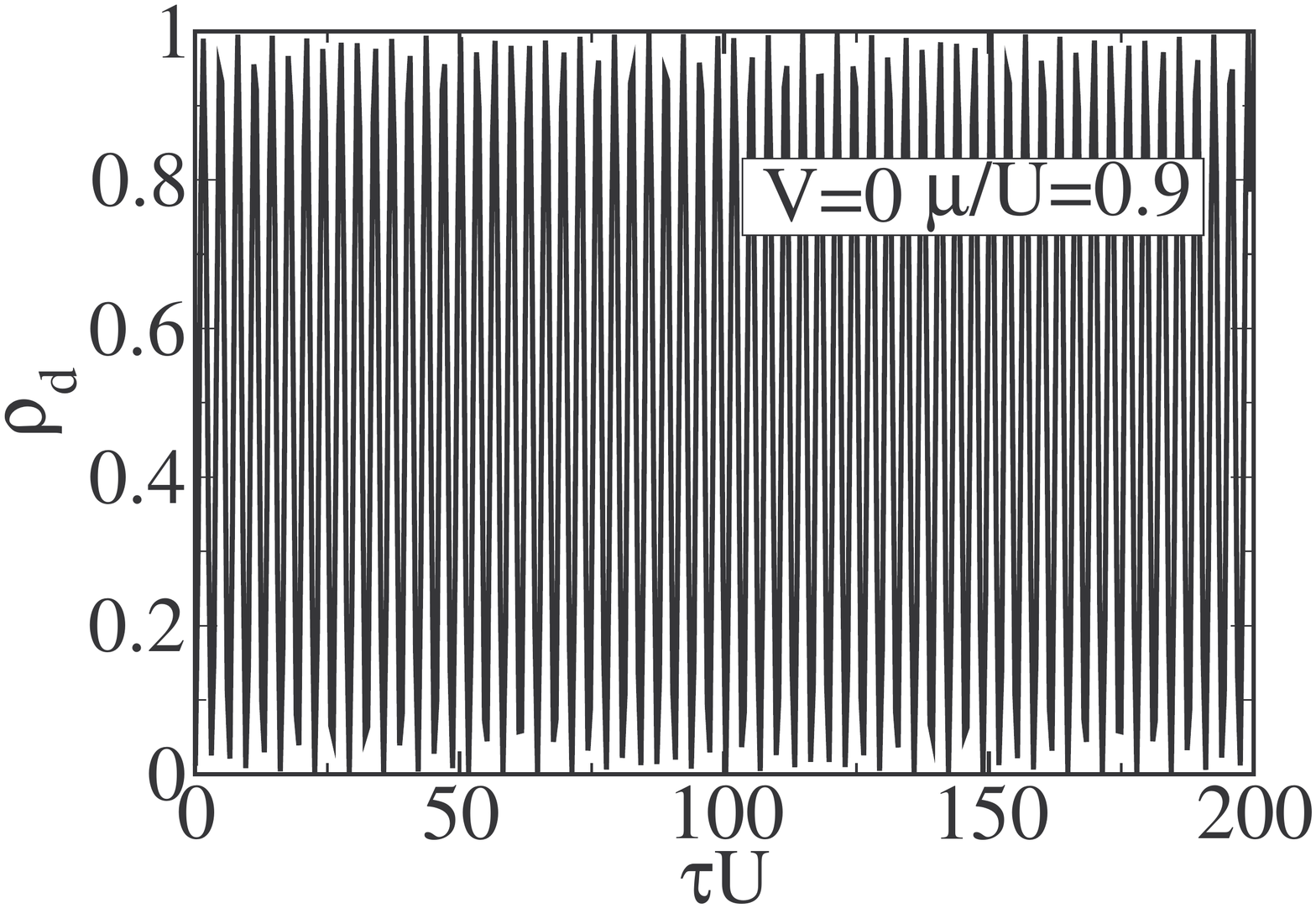}
\includegraphics[width = 0.38\columnwidth,angle=270]{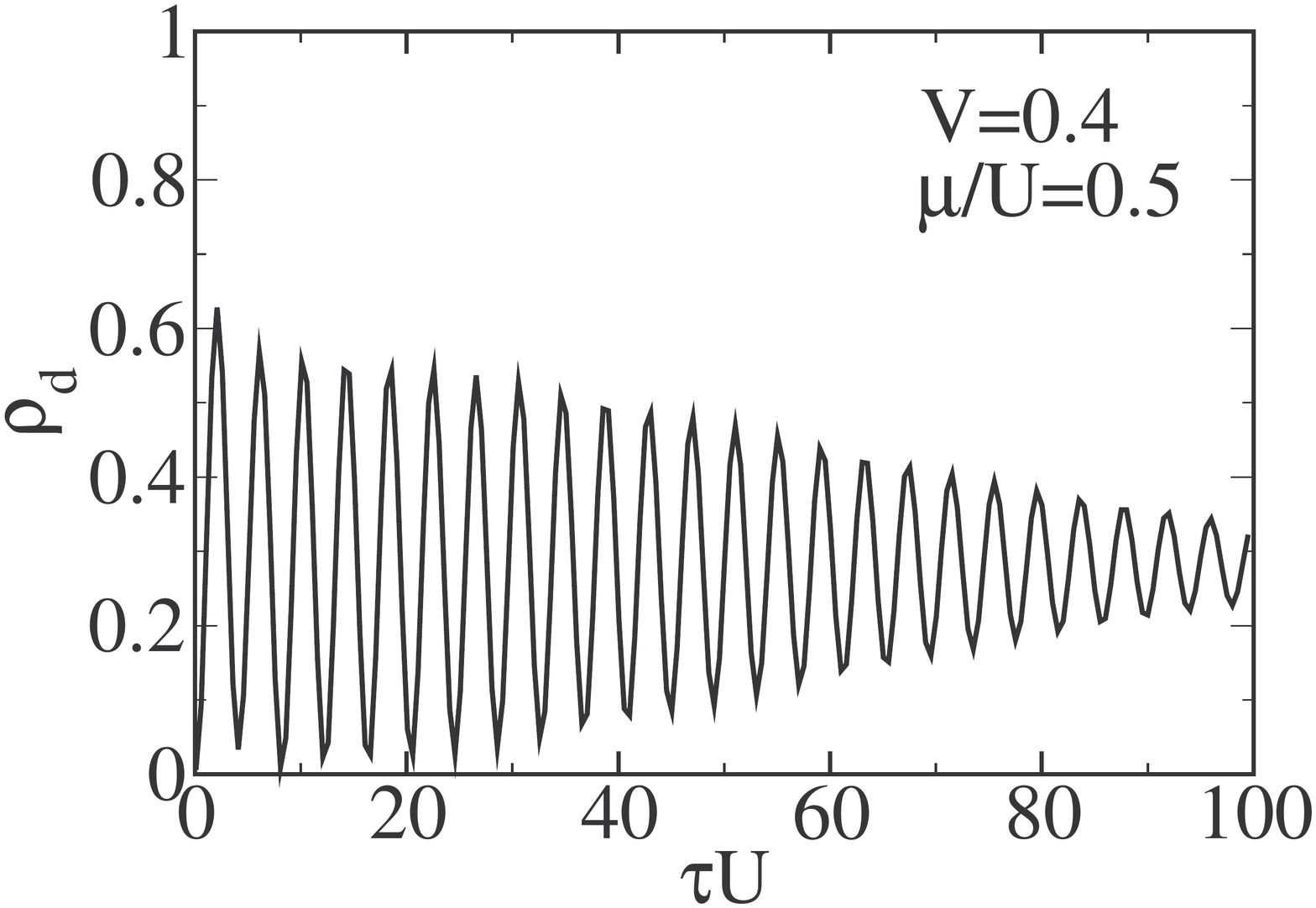}
\includegraphics[width = 0.38\columnwidth,angle=270]{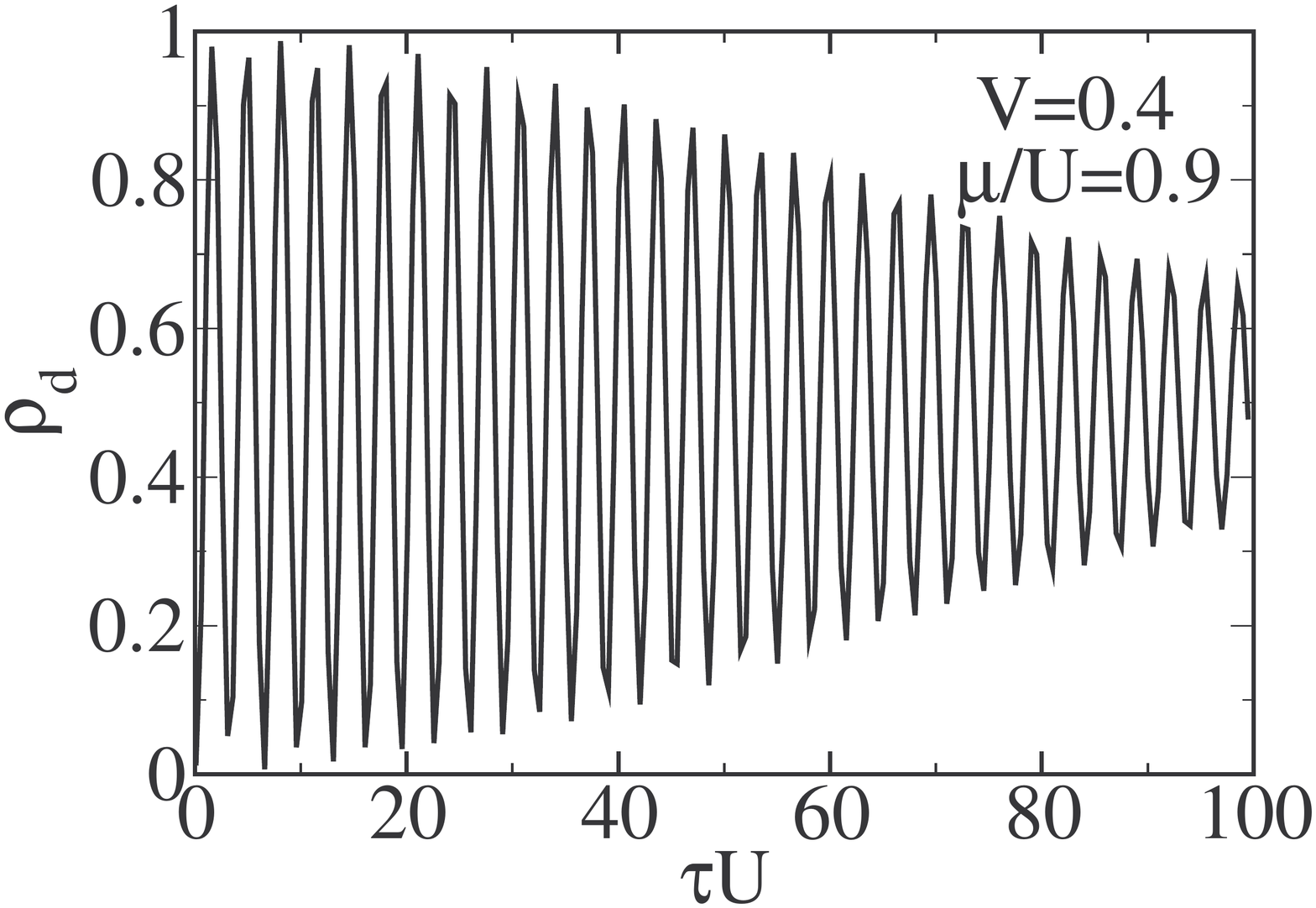}
\includegraphics[width = 0.38\columnwidth,angle=270]{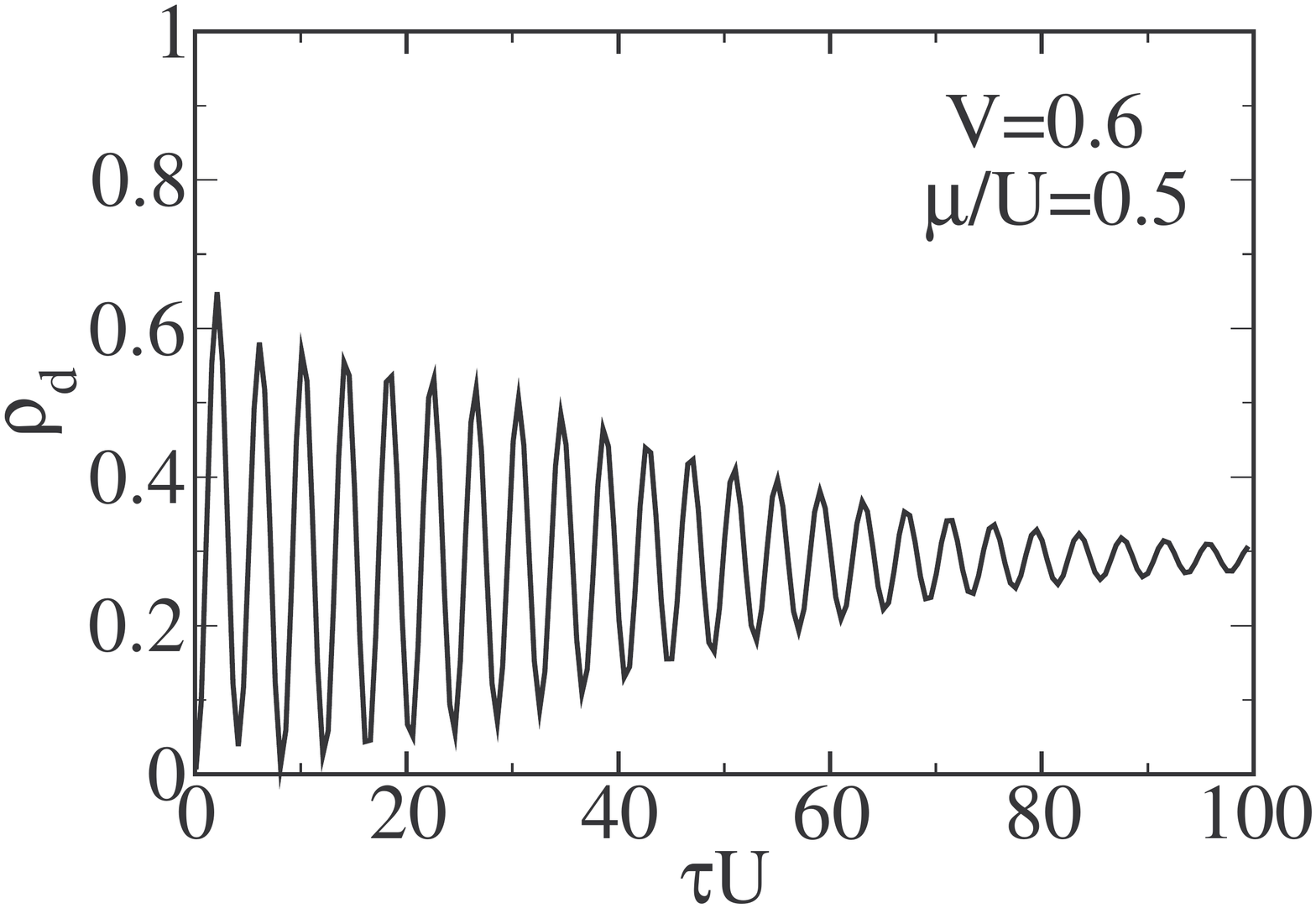}
\includegraphics[width = 0.38\columnwidth,angle=270]{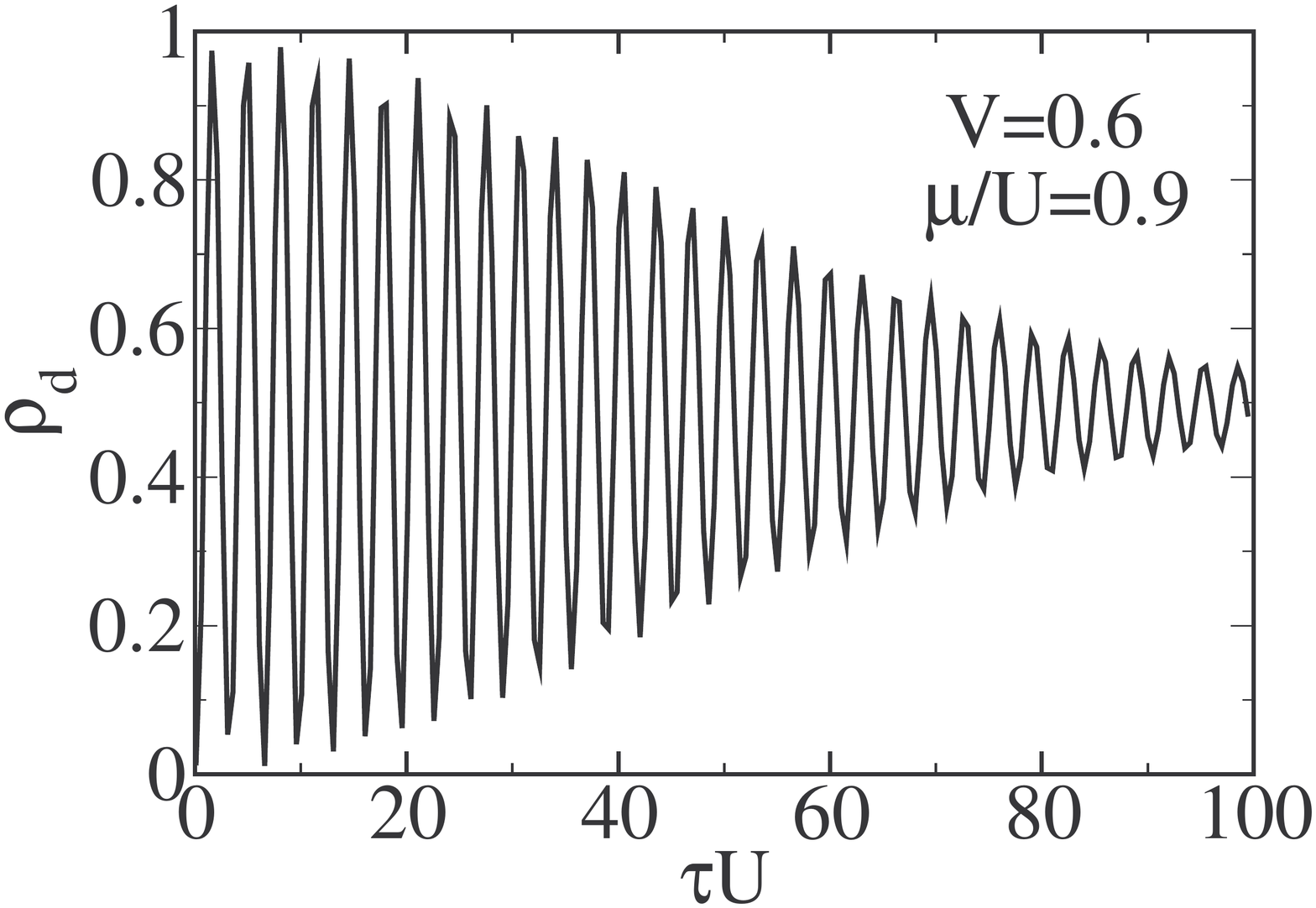}
\caption{Disordered averaged defect density as a function of rate of
change of hopping for: (a) top panel: clean case with $\mu/U=0.5$
(left) and $\mu/U=0.9$ (right) (b) middle panel: disorder potential
$V/U=0.4$ with $\mu/U=0.5$ (left) and $\mu/U=0.9$ (right) (c) lower
panel: disorder potential $V/U=0.6$ with $\mu/U=0.5$ (left) and
$\mu/U=0.9$ (right).} \label{fig:rhod}
\end{figure}

 Before concluding this section, we note that analogous phase
diagrams, which are in qualitative agreement with ours, have been
derived using single-site and multi-site mean-field mean-field
theories \cite{Krutitsky:NJP:06,Blakie:DBHMFT:PRA:07, Pisarski:PRA:11}, strong-coupling
expansions \cite{Freericks:PRB:96} and quantum Monte Carlo 
\cite{Pollet:MCDBH:PRL:09,Prokofyev:MCDBH:PRB:09,Ceperley:BG:PRB:11}.
All of these methods concur with ours regarding the qualitative
features of the phase diagram such as presence of a glassy region
between the superfluid and Mott phase in the presence of disorder
and the increase in the extent of this glassy region towards the
edge of the Mott lobes.

\section{Dynamics in the disordered Bose Hubbard Model\label{neqd}}

Our main goal in this paper is to study the dynamics of the disordered
Bose Hubbard model when a Hamiltonian parameter (in our case the
hopping $J$) is changed in time. Although a lot of work has been done
on the equilibrium phase diagram of the disordered Bose Hubbard model,
very little is known about the dynamics of this system. In this
context it is worth noting that the variational wavefunction and the
canonical transformation is especially well suited to treat the
dynamics in this system. For a dynamically changing system one can
write down a variational wavefunction of the form
\beq
|\psi(t)\rangle =e^{-i\can[J(t)]}|\psi_0(t)\rangle ~~~~
|\psi_0(t)\rangle =\prod_i\sum_n f_{ni}(t)|n\rangle_i
\eeq
where the canonical transformation is evaluated with the instantaneous
value of the parameter $J(t)$. Note that since our canonical transform
was based on the idea of eliminating terms in the Hamiltonian, which
connects states differing by a large energy $\sim U$, this would lead
to a coarse grained dynamics valid for timescales much larger than
$U^{-1}$. Further, since the canonical transform was not an expansion
around a particular state (like the Mott state), this can faithfully
capture the evolution of the excitations that are inevitably created
during time evolution.

The Schrodinger equation can then be written as
\beq
i\dot{|\psi_0\rangle}=(H^\ast-\dot{\can^\ast})|\psi_0\rangle
\eeq
where $\dot{\can^\ast}=e^{i\can}\dot{\can}e^{-i\can}$ and the initial ground state is evolved according to this equation.

At this point it is useful to look at the particular form of dynamics
we are interested in. We start our system in the ground state with an
initial value of $J_i$, which puts it in the superfluid phase. We then
decrease $J$ linearly to a very small value $J_f$ close to the atomic
limit with a rate $\tau^{-1}$. We then ramp back to our initial value
$J_i$ with the same ramp rate. The explicit time dependence of the
hopping parameter is given by
\bqa
\no J(t)& =&J_i+(J_f-J_i)\frac{t}{\tau}~~~~~  t<\tau\\
& = & J_f+(J_i-J_f)\frac{t}{\tau}~~~~~ t>\tau
\eqa
The effective Hamiltonian $H^\ast$ gives rise to energy scales of $J$,
$U$ and $J^2/U$, while the $\dot{S}$ term generates scales of $\Delta
J/(U\tau)$, $\Delta J J(t)/U^2\tau^2$ etc. We assume $U\tau >1$ (later
we will mostly be interested in the regime $U\tau \gg 1$), and we will
only keep the first order term $\dot{\can^1}$ in the dynamical
equations. This leads to a notable simplification; since
$\dot{i\can^1}\propto i\can^1$, $\dot{i\can^\ast}=\dot{i\can}$,
i.e. there is no Berry phase contribution from rotating the
$i\dot{\can}$ term. We note that this simplification goes away if we
include higher order terms in $i\can$.

\begin{figure}[t]
\includegraphics[width = 0.38\columnwidth,angle=270]{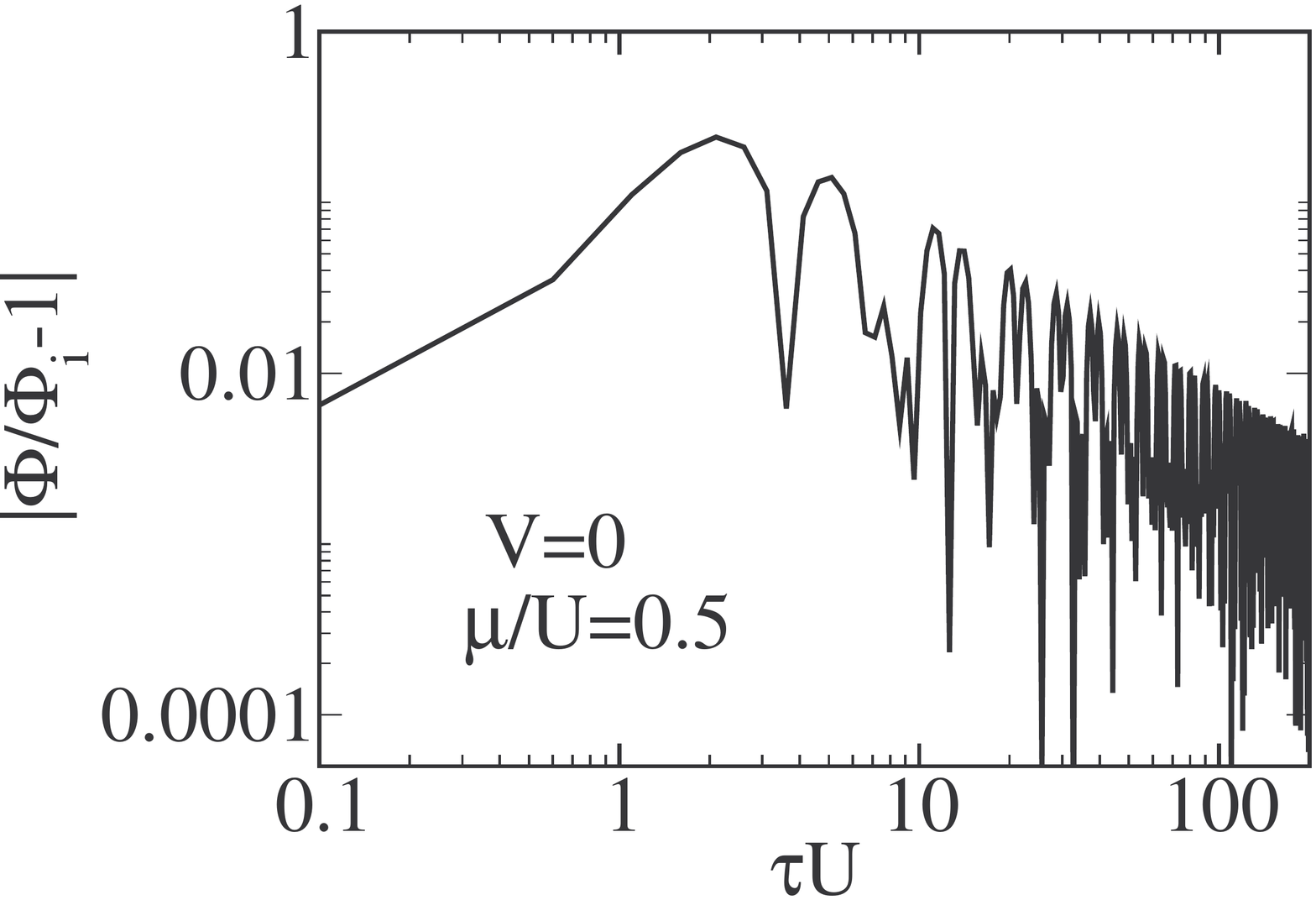}
\includegraphics[width = 0.38\columnwidth,angle=270]{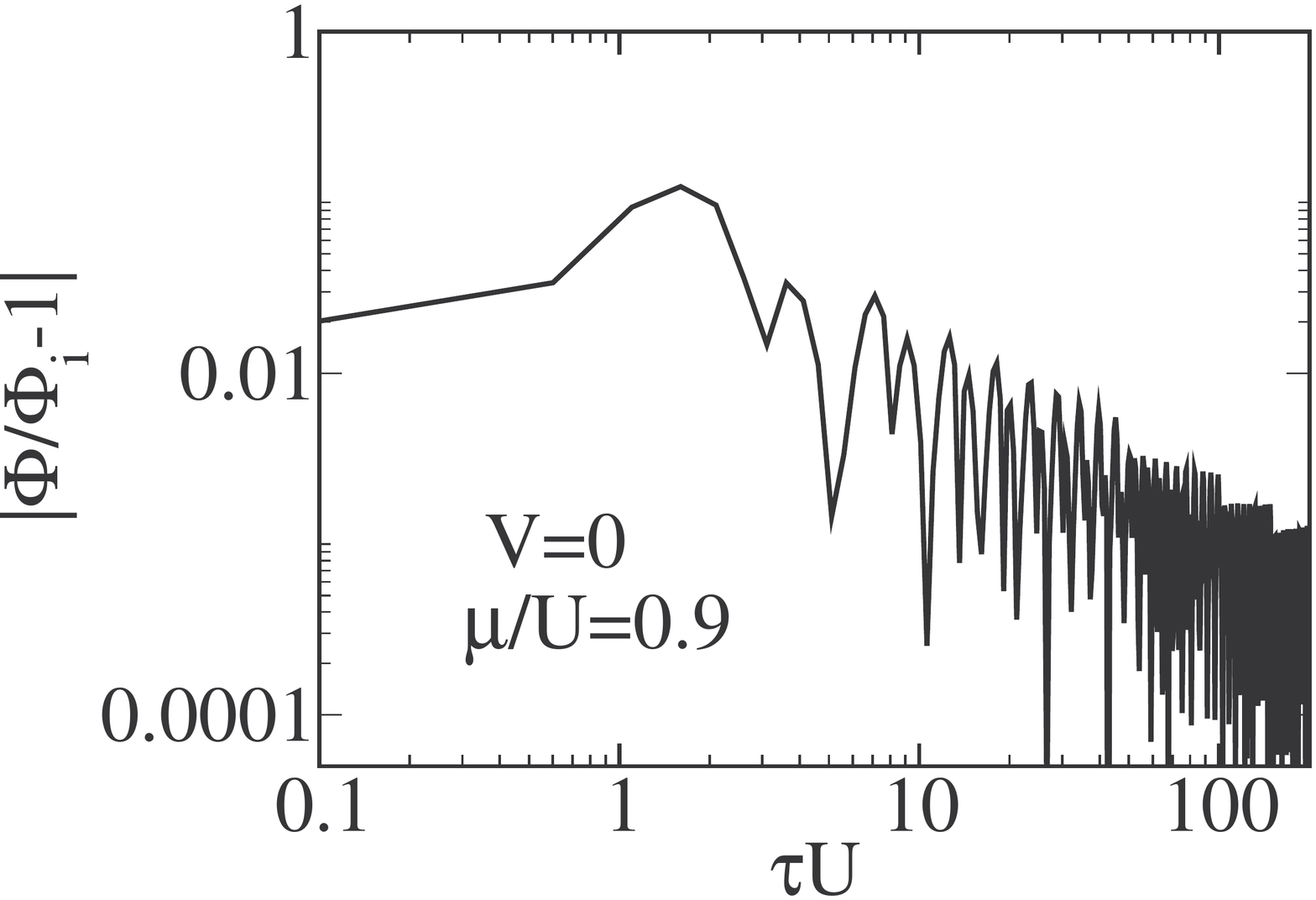}
\includegraphics[width = 0.38\columnwidth,angle=270]{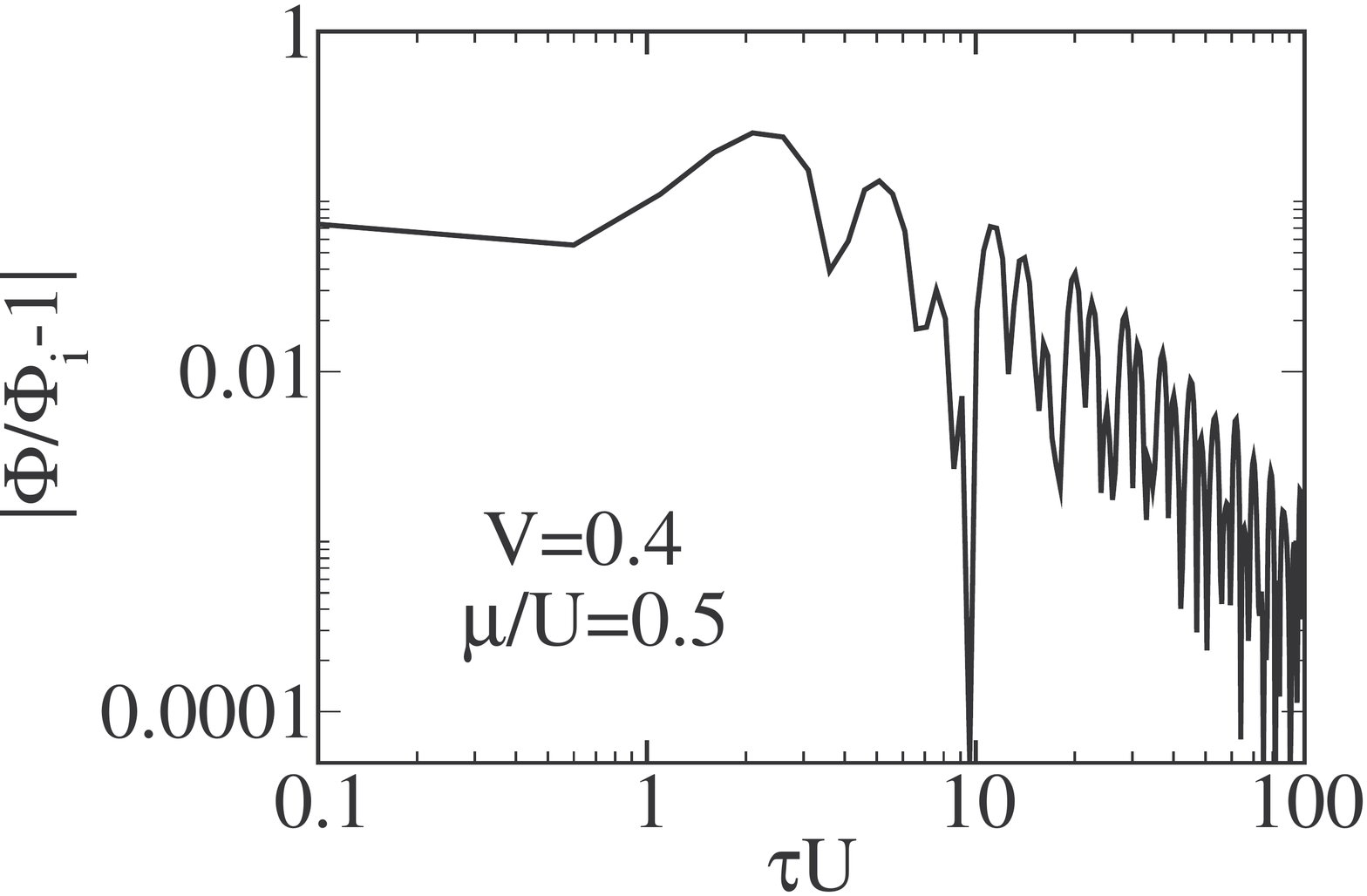}
\includegraphics[width = 0.38\columnwidth,angle=270]{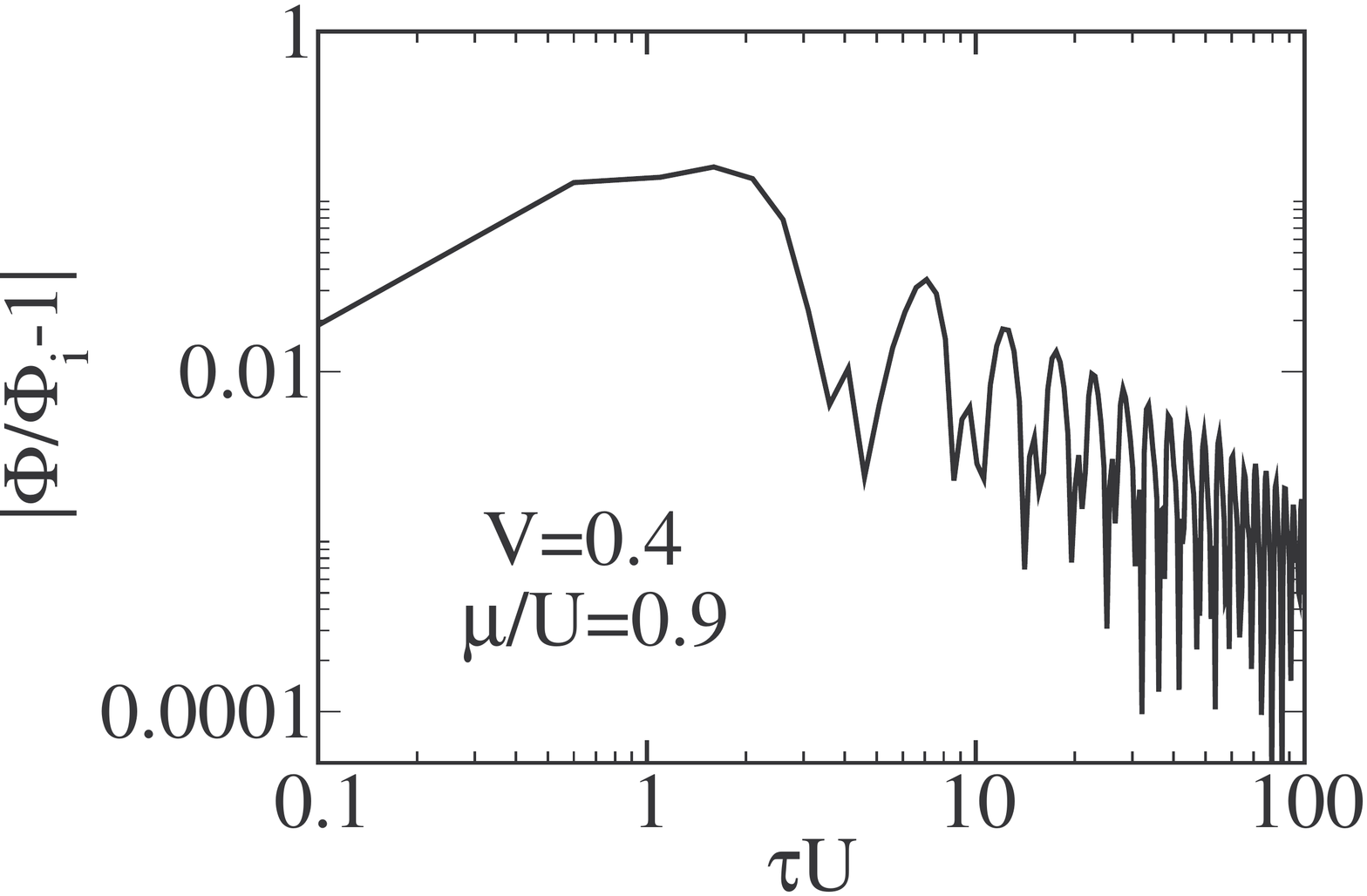}
\includegraphics[width = 0.38\columnwidth,angle=270]{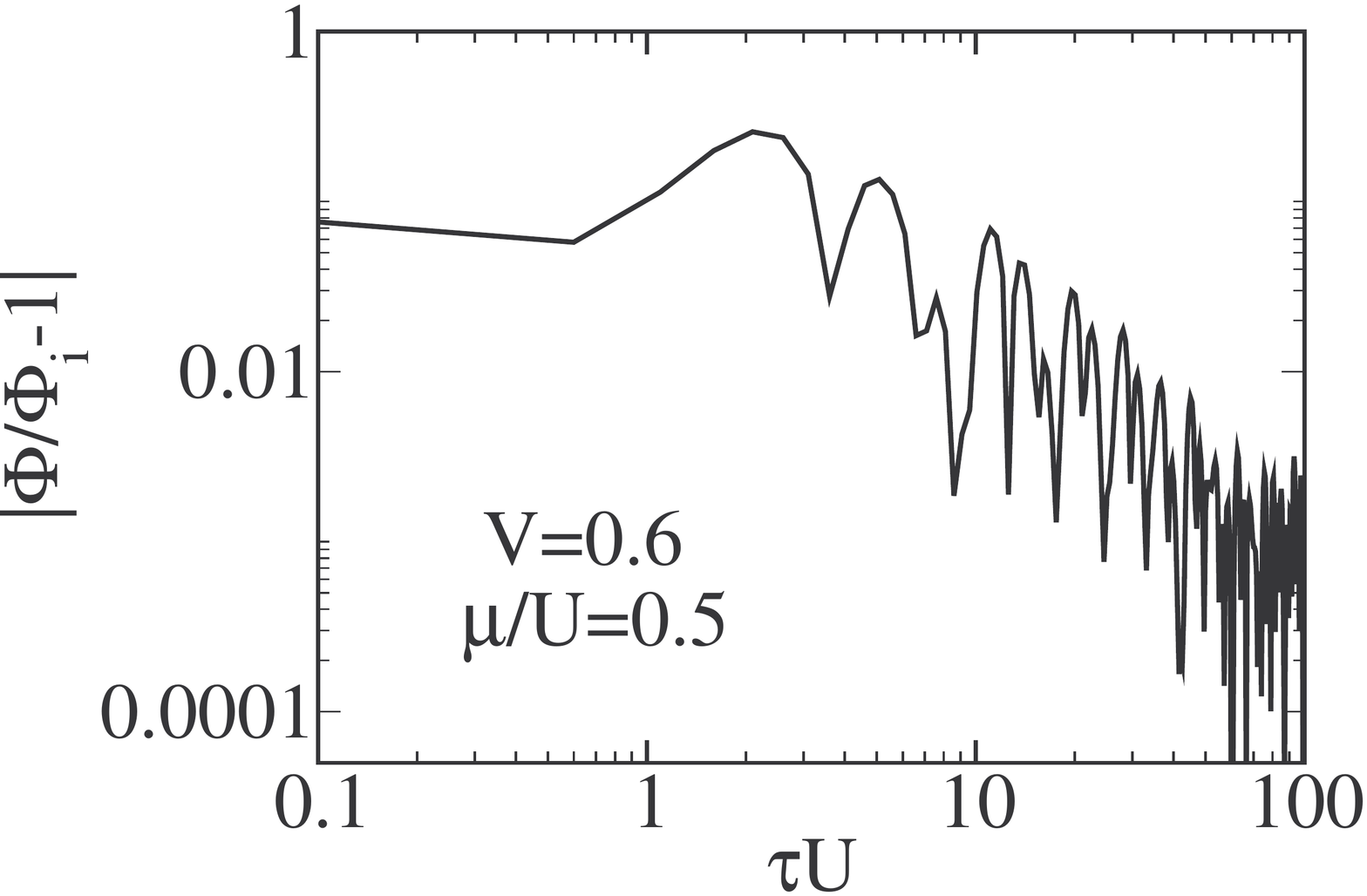}
\includegraphics[width = 0.38\columnwidth,angle=270]{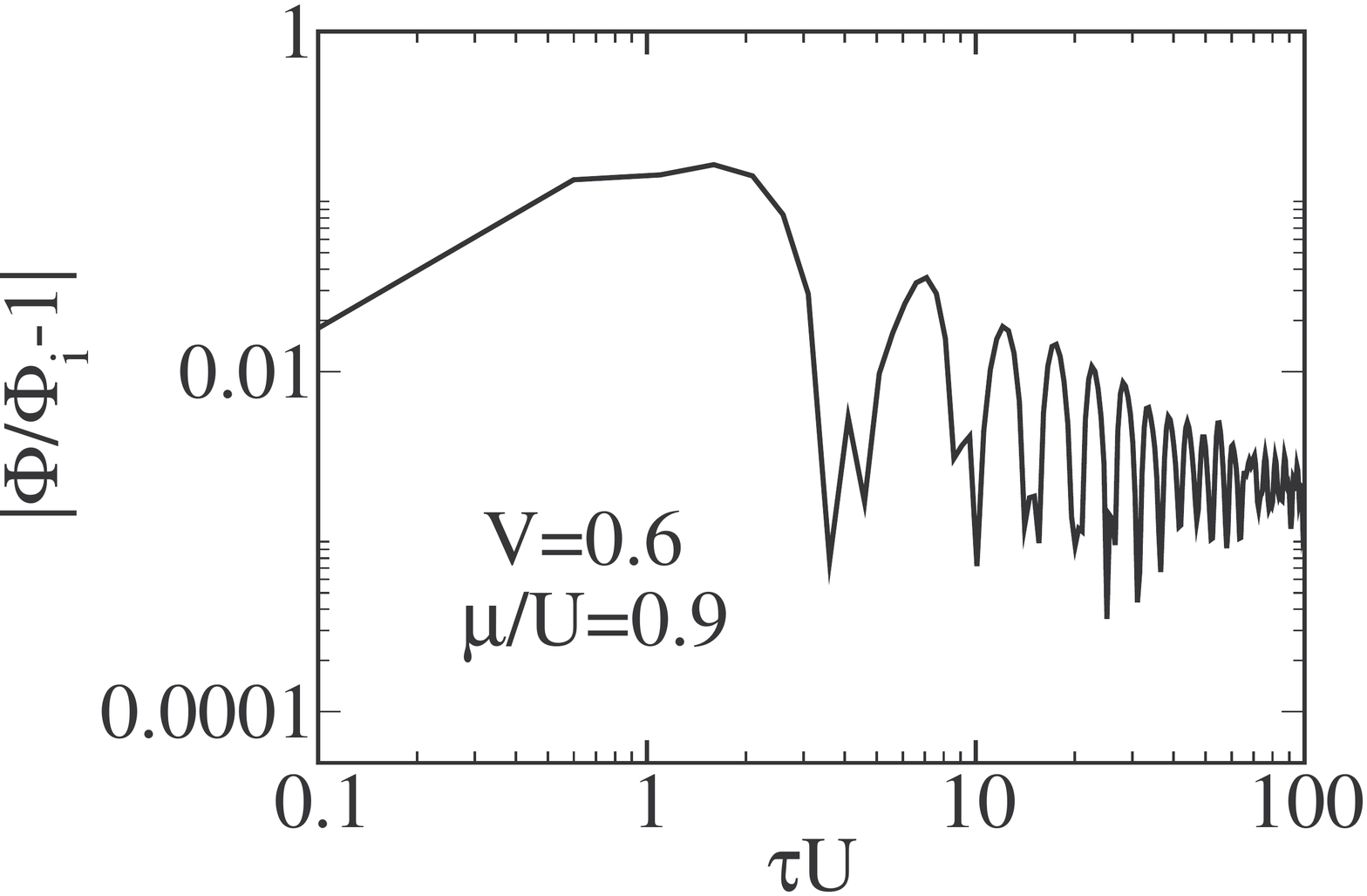}
\caption{Relaxation of superfluid order parameter as a function of
  rate of change of hopping. The deviation of $|\Phi(2\tau)/\Phi(0)|$
  from $1$ is plotted for: (a) top panel: clean case with $\mu/U=0.5$
  (left) and $\mu/U=0.9$ (right) (b) middle panel: disorder potential
  $V/U=0.4$ with $\mu/U=0.5$ (left) and $\mu/U=0.9$ (right) (c) lower
  panel: disorder potential $V/U=0.6$ with $\mu/U=0.5$ (left) and
  $\mu/U=0.9$ (right).}
\label{fig:phi}
\end{figure}
We are interested in the excitations created as we ramp down to the
atomic limit and ramp back up to the initial value of $J/U$. To study
this we look at the defect density which, for a given disorder
configuration is given by
\beq \rho_d(\tau)=\frac{1}{N_s}\sum_i 1-| \langle
\psi_0^i(2\tau)|\psi_0^i(0)\rangle|^2 \eeq
where $|\psi_0^i(t)\rangle =\sum_nf_{ni}(t)|n\rangle_i$ is the local
Gutzwiller wavefunction at time $t$. Note that since the final and
initial values of $J/U$ are same, the canonical transformation
operator does not affect this definition of defect density. For the
disordered Bose Hubbard model we study the defect density averaged
over many disorder realizations.  The defect density is plotted as a
function of the time constant $\tau$ for various values of $V/U$ and
$\mu/U$ in Fig~\ref{fig:rhod}. The top panel shows the clean case
($V=0$) results for (left): $\mu/U=0.5$, where one passes close to the
Mott lobe tip as one decreases the hopping $J$, and (right):
$\mu/U=0.9$, which is far away from the Mott tip. The defect density
shows oscillatory behaviour with $\tau$ with the large $\tau$ ($U\tau
\gg 1$) limit exhibiting an envelope which is constant with $\tau$. We
note that we have taken care to fix the gauge during the time
evolution and hence the oscillations are not a result of the system
sampling different gauge configurations in time. Rather, the return of
the system to its initial state ($\rho_d=0$) for certain rates of
change of hopping has similar origins as found in
Ref.~\onlinecite{Pekker:DYNFRZ:arx:12}, where the system was found to return to its
initial state under the influence of a periodic drive for certain
drive frequencies. The constant envelope characterizes the fact that
within the canonical transformation there is a low energy state
orthogonal to the initial ground state (with the degeneracy broken on
a scale of $\sim J^2/U$). The limiting constant value is $\sim 1$ away
from the Mott tip, where it is easy to create excitations and goes to
$\sim 0.6$ near the Mott tip. The middle panel shows the defect
density as a function of the ramp rate for a disordered system
characterized by $V/U=0.4$, while the lower panel shows a system with
$V/U=0.6$. The oscillatory behaviour persists, but presence of
disorder damps the oscillations, with the defect density showing an
exponential decay with the inverse ramp rate. Disorder leads to
scattering and lifts the degeneracy of the low lying state, thus
leading to an exponential decay of the defect density. It is also
clear by comparing the middle and the lower panel figures that as
$V/U$ increases, the damping timescale becomes smaller. In the large
$\tau$ limit, the defect density goes to a constant value, which is
almost independent of $V/U$ and depends crucially on $\mu/U$. For
$\mu/U =0.5$, this value is $\sim 0.3$, while for $\mu/U=0.9$, this
value increases to $\sim 0.6$. The finite value of the defect density
in the large $\tau$ limit is expected, as the system starts from
superfluid phase with associated gapless modes and hence there is no
excitation gap to protect defect creation in the slow ramping limit.

\begin{figure}[t]
\includegraphics[width = 0.38\columnwidth,angle=270]{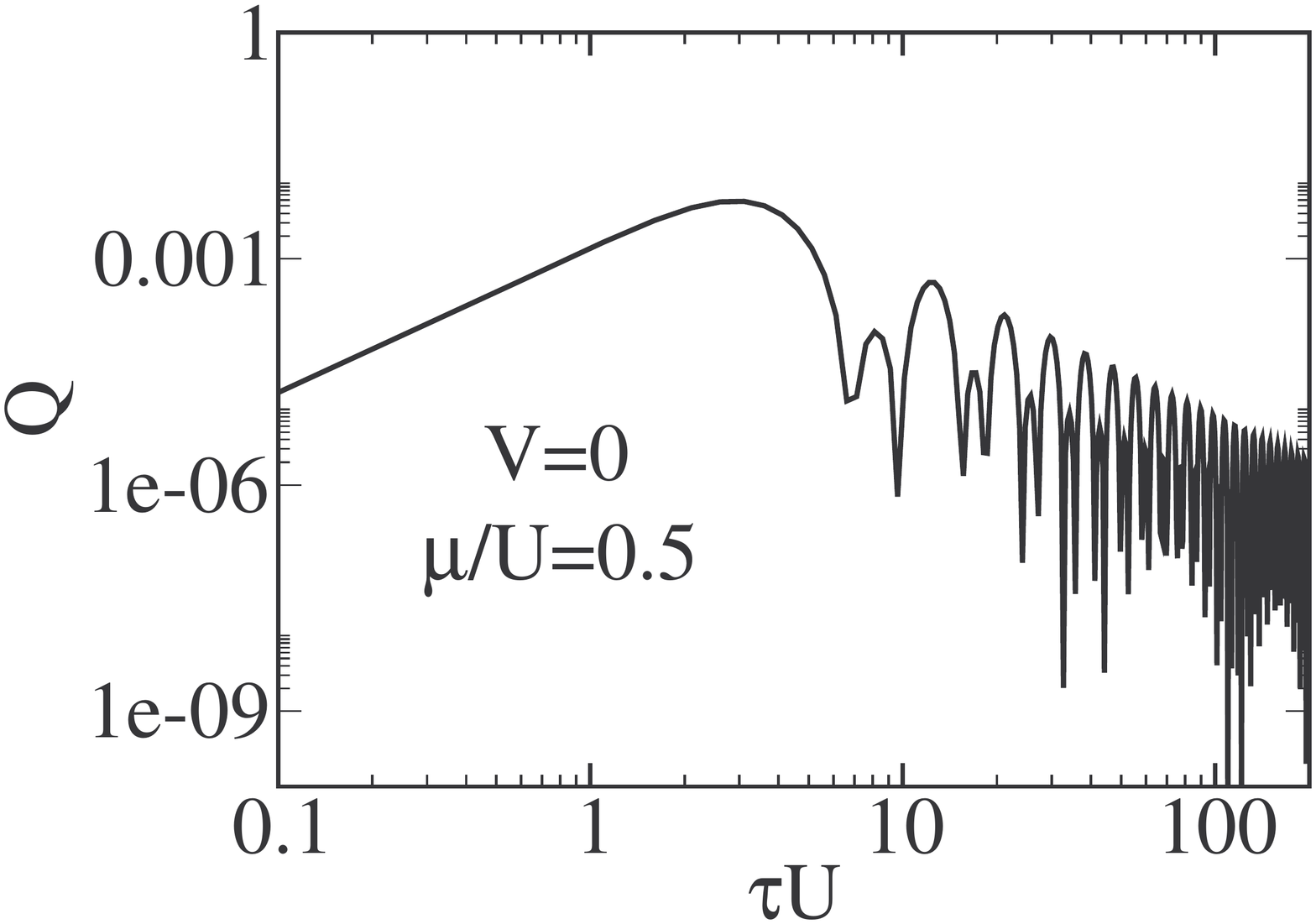}
\includegraphics[width = 0.38\columnwidth,angle=270]{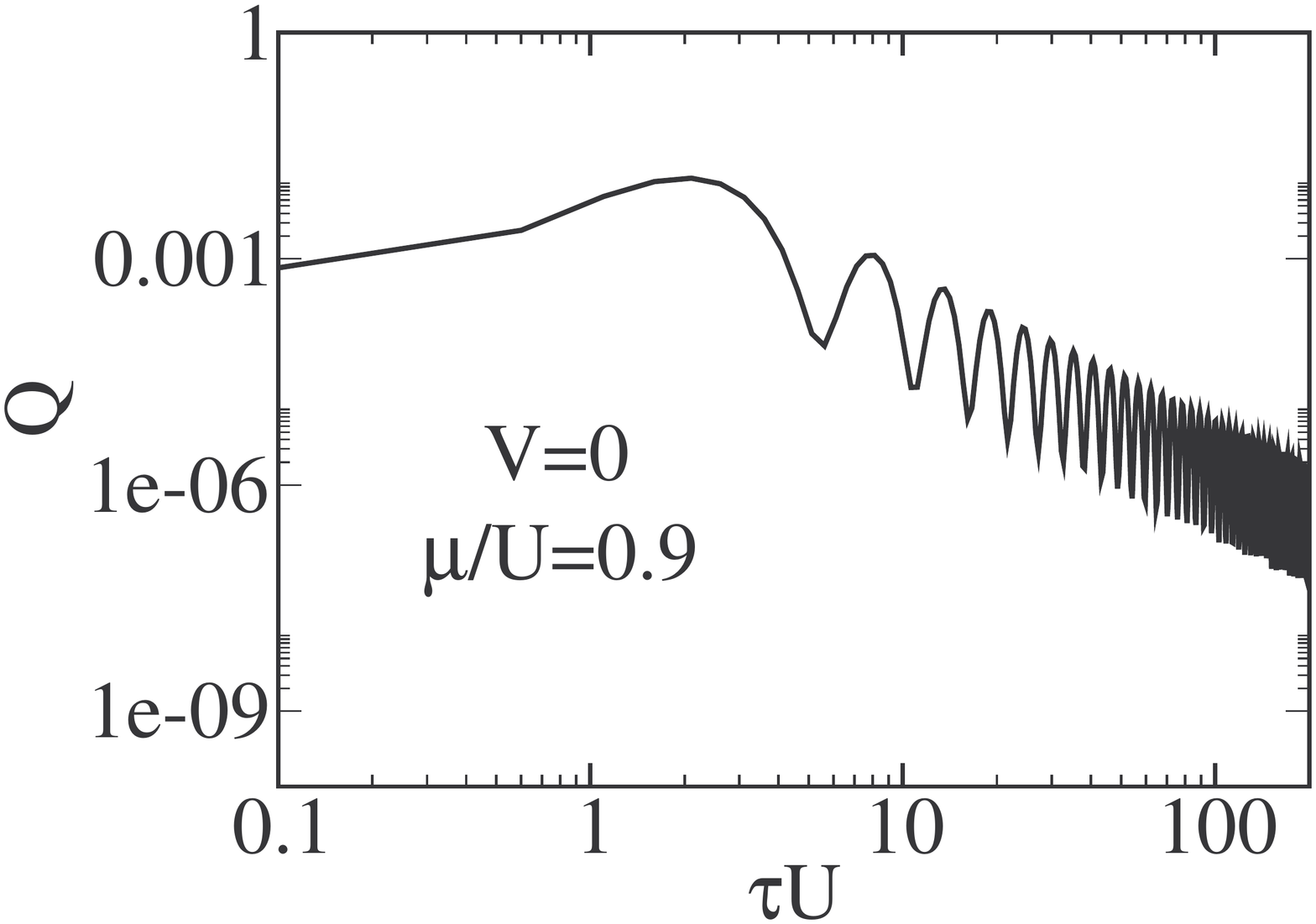}
\includegraphics[width = 0.38\columnwidth,angle=270]{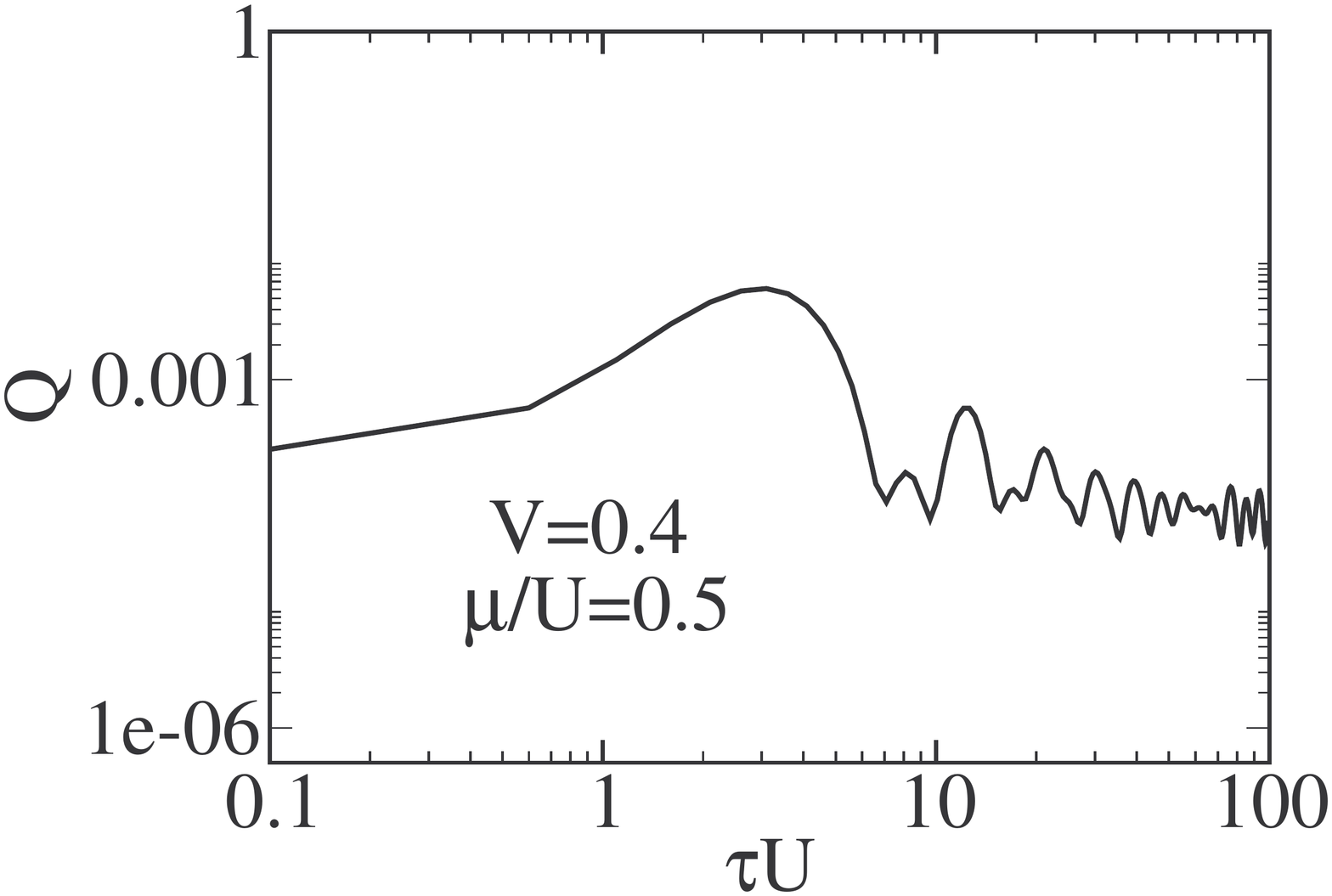}
\includegraphics[width = 0.38\columnwidth,angle=270]{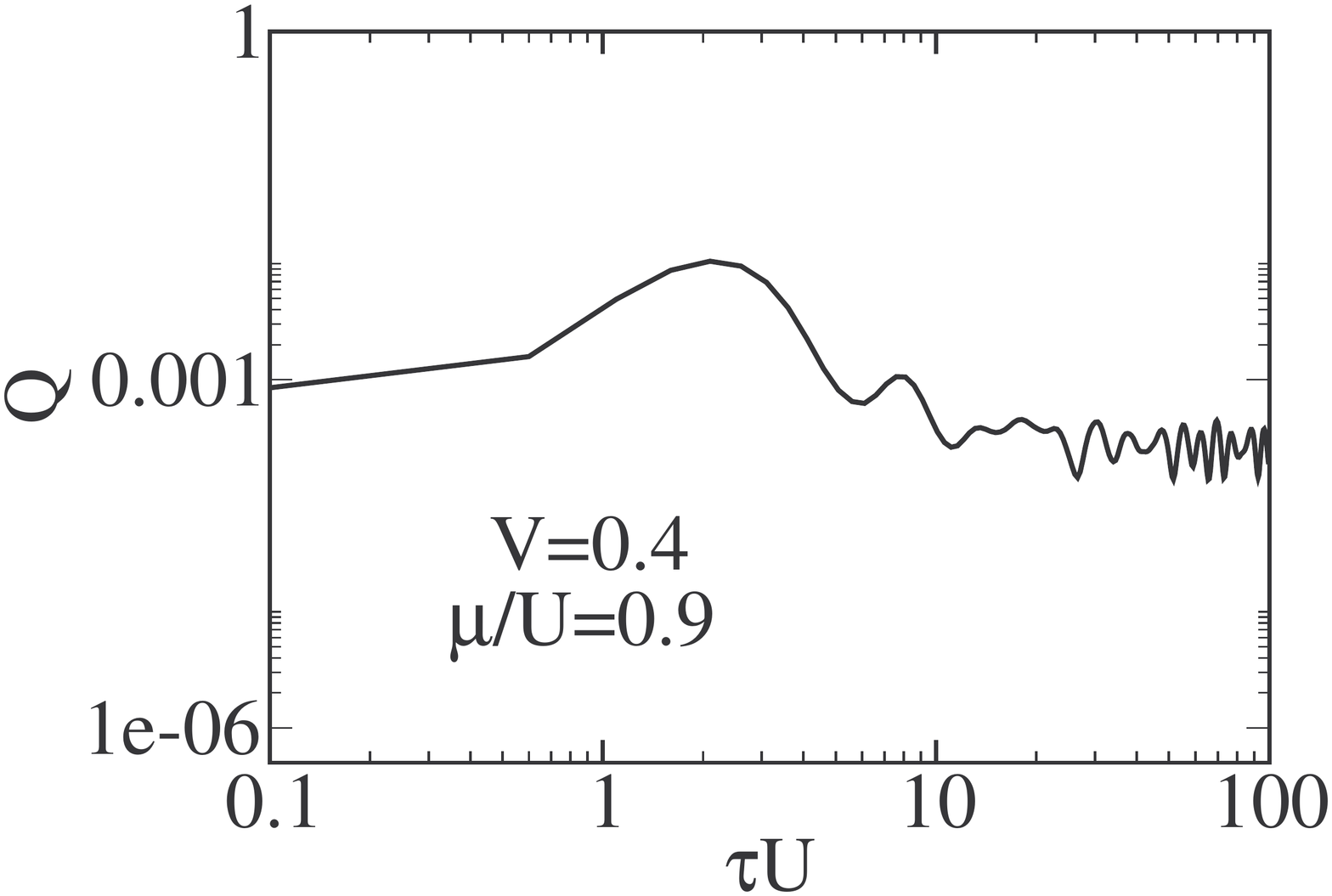}
\caption{Residual energy in the system as a function of
  rate of change of hopping for: (a) top panel: clean case with $\mu/U=0.5$
  (left) and $\mu/U=0.9$ (right) (b) bottom panel: disorder potential
  $V/U=0.4$ with $\mu/U=0.5$ (left) and $\mu/U=0.9$ (right).}
\label{fig:Q}
\end{figure}

We have also studied the evolution of the superfluid
order parameter
\beq
\Phi(t)=\frac{1}{N_s}\sum_i\langle \psi(t)| b_i|\psi(t)\rangle
\eeq
as the hopping is ramped down and up. To ensure normalization it is
easiest to look at the ratio $r=\Phi(2\tau)/\Phi(0)$ and then construct
the disorder average of this quantity. The disorder averaged $r$ goes
to $1$ in the large $\tau$ limit and hence we look at $|r|-1$ to
determine how the order parameter relaxes to its initial value as a
function of the ramp rate. This quantity is plotted in
Fig.~\ref{fig:phi} for the clean case (top panel) and the disordered
case for $V/U=0.4$ (middle panel) and $V/U=0.6$ (lower panel). In the
clean case, the quantity $|r|-1$ clearly shows a power-law scaling
with an asymptotic $1/\tau$ envelope on top of oscillations in the
large $\tau$ limit. The relaxation in the disordered case is not so
simple. Although $|r|-1$ goes down with $\tau$, we have not been able
to clearly extract either a power law or an exponential scaling from
the large $\tau$ limit. There is a substantial window of $\tau$ values, where a $1/\tau$ power law can be defined, but the data seems to deviate from this scaling for larger values of $\tau$.

We also study the residual energy pumped into the system in the process of
ramping down the hopping and returning back to the original
value. This is important as there is a lack of in-situ measurements of
temperature in optical lattices and the energy pumped into the system
is often taken as a bound on the amount of heating in the system. The
excess energy of the system in the final state is given by
\beq
Q=\langle \psi(2\tau)|H|\psi(2\tau)\rangle-\langle \psi(0)|H|\psi(0)\rangle
\eeq
The excess energy of the system as a function of ramp rate is shown in
Fig.~\ref{fig:Q}. In the clean case, the energy decays with $\tau$,
with a $1/\tau^2$ envelope in the large $\tau$ limit. One way to
understand this is that within a Gross-Pitaevsky description, the
lowest order dependence of the energy on the order parameter is $E\sim
|\Phi|^2$, and so, a $1/\tau$ relaxation of the order parameter leads
to a $1/\tau^2$ energy relaxation in the system. In the disordered case, although the excess energy decreases with $\tau$, numerical accuracy of the data forbids a clear extraction of an asymptotic limit.

\section{Conclusions\label{concl}}

In this paper, we have studied the equilibrium and non-equilibrium
properties of the disordered Bose Hubbard model using a new
variational wavefunction approach. Our variational wavefunction
implements the canonical Schrieffer-Wolf transformation, which has been
extensively used for strongly interacting Fermions to the case of
strongly repulsive Bosons in a non-uniform potential background. We
have determined the equilibrium phase diagram of the system, which
shows the expected Mott insulator, Bose glass and the superfluid
phases. Our phase diagram is qualitatively similar to the phase
diagram obtained by more sophisticated techniques.

We have also studied the non-equilibrium properties of the disordered
bosons within this variational approach. We have focused on the
specific dynamic process, where, starting from the system in its
ground state in the superfluid phase, the hopping parameter $J$ is
ramped down linearly with a rate $\tau^{-1}$ to a value very close to
$0$. The hopping is then ramped back with the same rate to its initial
value. We look at the response of the system to this non-equilibrium
cyclic process by studying the density of excitations, the superfluid
order parameter and the energy of the excitations in the final state
obtained from the time evolution of the initial ground state. In the
clean system ($V=0$), we find that all the three quantities show
oscillatory behaviour as a function of the inverse ramp rate
$\tau$. The asymptotic envelope of the defect density is a constant,
while the order parameter and the energy decays as $1/\tau$ and
$1/\tau^2$ respectively in the large $\tau$ ($U\tau \gg 1$) limit. In
the disordered system, the oscillations persist, although they are
damped by disorder scattering. The defect density, as a function of
the inverse ramp rate, oscillates with an exponentially decaying
envelope. The decay of the defect density as a function of $\tau$
increases with increasing $V/U$, while the value of the excitation
density in the large $\tau\rightarrow \infty$ limit is independent of
$V/U$, but depends on the value of $\mu/U$, i.e. it increases as one
moves away from the tip of the Mott lobe. The deviation of the
superfluid order parameter from its initial value as well as the
energy pumped into the system decays with decreasing ramp rate, but
numerical noise prohibits a clear extraction of a large $\tau$
asymptotic scaling.

The variational wavefunction and the associated canonical transform
used in this paper provides a new analytic way of treating the problem
of disordered strongly interacting bosons on a lattice. In fact, this
technique can be used to study any one body potential (including
harmonic trap potentials relevant to the cold atom experiments). In
its current form, the variational wavefunctions capture the essential
physics of the superfluid, Bose glass and Mott insulator phases in the
low disorder limit $V/U \ll1$. Although we have stretched the
technique to $V/U \sim 0.6$, the numerical accuracy of the method
decreases and quantitative match of the phase diagram with the Monte
Carlo results deteriorates. A more complete formulation, which is beyond
the scope of this paper, would incorporate the fact that for
$U>|v_i-v_j|\gg J$, the hopping on the bond between $i$ and $j$ is
completely frozen, while for $v_i-v_j \sim nU$, there is a low energy
hopping process which changes the number of multiple occupancies in
the system. Further development along these lines would lead to higher
numerical accuracy and wider applicability of this new technique.

\medskip

This work is supported by AFOSR JQI-MURI, ARO-DARPA-OLE, and NSF-JQI-PFC.

\bibliography{bgdyn}

\end{document}